\documentclass[twocolumn]{aastex62}

\usepackage{wrapfig} 
\usepackage{savesym}
\savesymbol{tablenum} 
\usepackage{siunitx}
\usepackage{multirow}
\usepackage[caption=false]{subfig}
\graphicspath{{./}{figures/}}

\accepted{January 31, 2022}

\shorttitle{Ly$\alpha$ and \CII\ from $z\approx6$ QSOs}
\shortauthors{Drake et al.}

\def\poog{P$009\!-\!10$} 
\def\wolf{P$359\!-\!06$} 
\def\psixfive{P$065\!-\!26$} %
\def\pmerg{P$308\!-\!21$} 
\def\pthreesix{P$036\!+\!03$} %
\def\pchiara{P$231\!-\!20$} 
\def\pthreetothree{P$323\!+\!12$} %
\def\jema{J$0305\!-\!3150$} 

\def\Roche{\mbox{J2228$+$0110}} 

\def\CFHQS{\mbox{J2329$-$0301}} 

\def\nosample{18} 
\def\nosubsample{8} 
\def\sigmahalo{2} 
\def\kernelEma{$\sigma_{spat}=0.2$\,\arcsec and $\sigma_{spec}=2.5$\,\AA} 

\def\pkpcatsix{5.71} 

\def\requiem{{\sc{requiem}}} 

\def\PALyathreesix{$\approx\,270^{\circ}$}
\def\PALyachiara{$\approx\,315^{\circ}$}

\def\SBunits{erg s$^{-1}$ cm$^{-2}$ arcsec$^{-2}$}
\def\kms{km s$^{-1}$}

\def\CII{[C{\sc{ii}]}}

\def\MUSE{{\sc{muse}}}
\def\ALMA{{\sc{alma}}}

\begin{document}

\title{The Decoupled Kinematics of high-z QSO Host Galaxies and their Ly$\alpha$ halos}

\correspondingauthor{Alyssa Drake}
\email{drake@mpia.de}

\author[0000-0002-0174-3362]{Alyssa B. Drake}
\affil{Max Planck Institute f{\"u}r Astronomie, K{\"o}nigstuhl, Heidelberg, Germany}
\affil{Centre for Astrophysics Research, Department of Physics, Astronomy and Mathematics, University of Hertfordshire, Hatfield AL10 9AB, UK}

\author[0000-0002-9838-8191]{Marcel Neeleman}
\affil{Max Planck Institute f{\"u}r Astronomie, K{\"o}nigstuhl, Heidelberg, Germany}

\author[0000-0001-9024-8322]{Bram P.\ Venemans}
\affil{Max Planck Institute f{\"u}r Astronomie, K{\"o}nigstuhl, Heidelberg, Germany}

\author[0000-0001-8695-825X]{Mladen Novak}
\affil{Max Planck Institute f{\"u}r Astronomie, K{\"o}nigstuhl, Heidelberg, Germany}

\author[0000-0003-4793-7880]{Fabian Walter}
\affil{Max Planck Institute f{\"u}r Astronomie, K{\"o}nigstuhl, Heidelberg, Germany}

\author[0000-0002-2931-7824]{Eduardo Ba{\~n}ados}
\affil{Max Planck Institute f{\"u}r Astronomie, K{\"o}nigstuhl, Heidelberg, Germany}

\author[0000-0002-2662-8803]{Roberto Decarli}
\affil{INAF - Osservatorio Astronomico di Bologna, Via Piero Gobetti, 93/3, 40129 Bologna BO, Italy}

\author[0000-0002-6822-2254]{Emanuele Paolo Farina}
\affil{Max Planck Institute for Astrophysics, Karl-Schwarzschild-Str, Garching, Germany}

\author[0000-0002-5941-5214]{Chiara Mazzucchelli}
\affil{European Southern Observatory, Alonso de Cordova 3107, Vitacura, Region Metropolitana, Chile}

\author[0000-0002-6849-5375]{Maxime Trebitsch}
\affil{Max Planck Institute f{\"u}r Astronomie, K{\"o}nigstuhl, Heidelberg, Germany}

\begin{abstract}

\noindent We present a comparison of the interstellar medium traced by [CII] (\ALMA), and ionised halo gas traced by Ly$\alpha$ (\MUSE), in and around QSO host galaxies at $z\sim6$. To date, \nosample\ QSOs at this redshift have been studied with both \MUSE\ and high--resolution \ALMA\  imaging; of these, \nosubsample\ objects display a Ly$\alpha$ halo. Using datacubes matched in velocity resolution, we compare and contrast the spatial and kinematic information of the Ly$\alpha$ halos and the host galaxies' \CII\ (and dust-continuum) emission. We find that the Ly$\alpha$ halos extend typically $3-30$ times beyond the interstellar medium of the host galaxies. The majority of the Ly$\alpha$ halos do not show ordered motion in their velocity fields, whereas most of the \CII\ velocity fields do. In those cases where a velocity gradient can be measured in Ly$\alpha$, the kinematics do not align with those derived from the \CII\ emission. This implies that the Ly$\alpha$ emission is not tracing the outskirts of a large rotating disk  that is a simple extension of the central galaxy seen in \CII\ emission. It rather suggests that the kinematics of the halo gas are decoupled from those of the central galaxy. Given the scattering nature of Ly$\alpha$, these results need to be confirmed with JWST IFU observations that can constrain the halo kinematics further using the non--resonant H$\alpha$ line.\\
\end{abstract}


\section{Introduction} \label{sec:intro}

High-redshift quasars (QSOs) present a unique opportunity to study some of the most extreme objects in the Universe, back to $<1$ Gyr after the Big Bang. In these objects vast amounts of material are funnelled onto galaxies' central supermassive black holes. This leads to accretion disks that shine with sufficiently high luminosities to allow their detection well into the epoch of reionisation (e.g. \citealt{Mortlock2011}, \citealt{Banados2018}, \citealt{Yang2020}). At $z\sim6$, several hundred QSOs have now been detected (e.g. \citealt{Banados2016}, \citealt{Jiang2016},  \citealt{Mazzucchelli2017b}) and it was recently discovered that some of  these QSOs are surrounded by giant Ly$\alpha$ halos. \cite{Drake2019} for instance presented deep  \MUSE\ observations of 5 QSOs at $z\sim6$, and revealed that 4 of these objects displayed Ly$\alpha$ halos comparable in extent and luminosity to their lower redshift counterparts at $2\!\leq\!z\!\leq\!3$ (e.g. \citealt{Borisova2016}, \citealt{ArrigoniBattaia2018}). This confirmed earlier evidence of Ly$\alpha$ halos at $z\sim6$ based on long--slit spectroscopy and/or narrow--band imaging for both the radio--loud QSO \Roche\ at $z=5.903$ (\citealt{Roche2014}, \citealt{Zeimann2011}), and \CFHQS\ at $z=6.43$ (\citealt{Goto2009}, \citealt{Willott2011}, \citealt{Goto2012} and \citealt{Momose2018}) and was later followed--up by \cite{Farina2019} who presented additional Ly$\alpha$ halos as part of the \MUSE\ snapshot survey ``\requiem".

\begin{figure*}
\figurenum{1}
\plotone{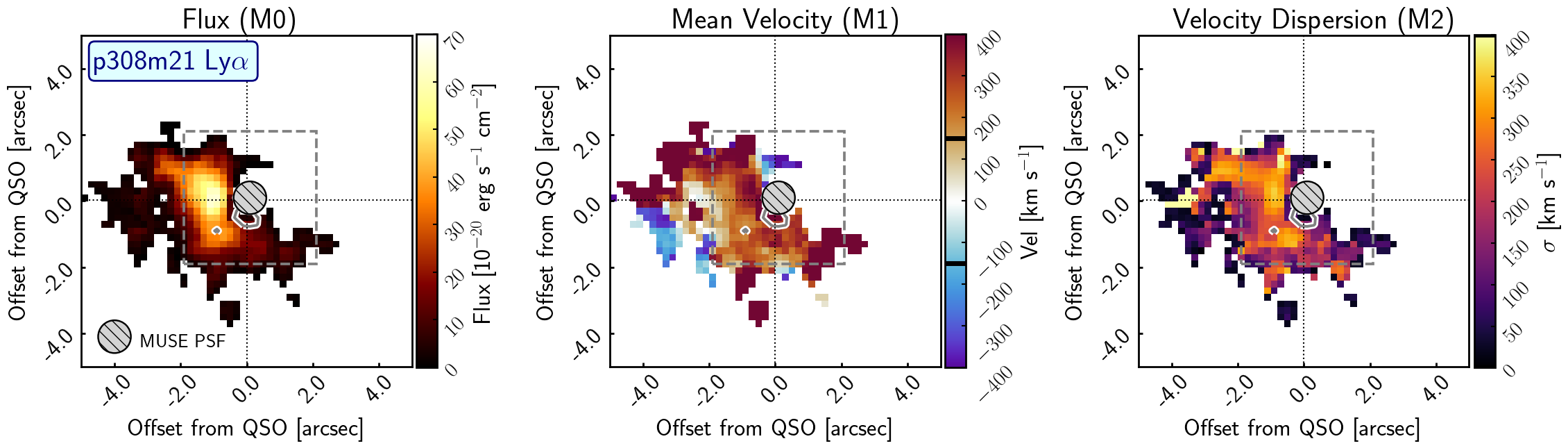}
\plotone{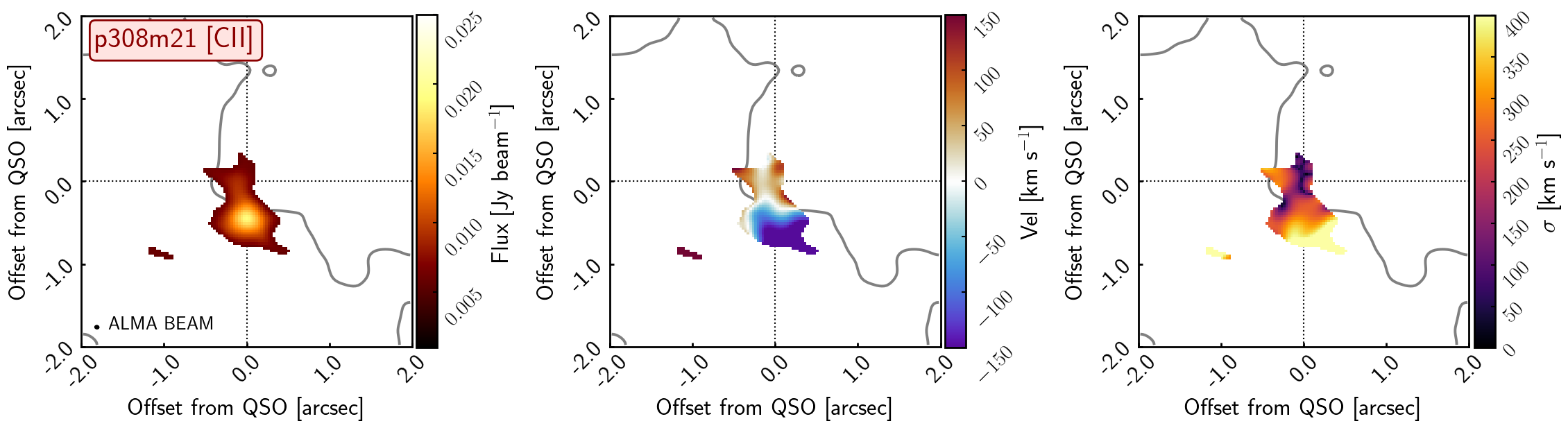}
\label{Fig: mom ex}
\caption{Moment maps for the Ly$\alpha$ halo (upper panels) and \CII\ emission (lower panels) for \pmerg. The upper panels are 10\arcsec\ a side. The area corresponding to complex residuals following PSF-subtraction is shown by the hatched circle. In addition, grey contours show the region from which \CII\ emission is mapped. The lower panels zoom-in to a box of 4\arcsec\ a side (grey dashed boxes in the top row), and display the \CII\ moment maps together with a grey contour that traces the edge of the Ly$\alpha$ halo shown above). The left-hand column shows the flux images (moment 0), the central column shows the velocity offset relative to systemic z (moment 1; note that the colourbar for the lower panels is a zoom-in to the range shown by two black bars on the colourbar in the upper panels) and the right-hand column shows the velocity dispersion ($\sigma$, moment 2) of the gas. Moment maps for the full sample can be found in the figure set available in the online journal (or in the Appendix on arXiv).}
\end{figure*}

Contemporaneously, observations from \ALMA\ have revealed the far--infrared (FIR) dust--continuum and \CII\ emission from the interstellar medium (ISM) in $\geq 27$ $z\sim6$ quasar host galaxies \citep{Decarli2018}. These observations confirmed earlier studies in that these objects are rich in gas and dust. Higher spatial--resolution observations from \ALMA\ reaching resolution of $\sim1$\,pkpc, have been analysed in a series of papers: These observations revealed centrally--concentrated dust emission around the supermassive black holes (traced by FIR continuum emission; \citealt{Venemans20}), and a diverse set of kinematics traced by \CII\ emission \citep{Neeleman2021} broadly split into three categories of equal number; disturbed, dispersion--dominated, or smoothly rotating. The third paper, \cite{Novak20}, determined that both \CII\ and dust continuum morphologies can be described with a two--component model consisting of a central steep component and an extended component of shallower gradient. While for the compact components the FIR continuum emission is more compact than the \CII, the extended components of both tracers extend over similar scales.

Despite the growing wealth of information now available for these objects, we have few constraints on the flow of gas into-- and around QSO host galaxies, and also the mechanisms governing the supply of pristine gas available to their rapidly--accreting supermassive black holes. Indeed it is a long--standing question for simulations how the cold gas which is funnelled along filaments connects to-- and feeds galaxies. Some numerical simulations have suggested that large Ly$\alpha$ halos will trace high angular-momentum gas as it is accreted (e.g \citealt{Stewart2011} and \citealt{Stewart2013}) --- observational evidence for this scenario is found in \cite{Prescott15}, where the authors report large-scale rotation of a collapsing gas structure. To date, no further rotation of gas in emission on this scale has been presented in the literature. It is prudent then, to compare and contrast available datasets, to gain insights on the relationship between the kinematics of the ISM {\em inside} QSO host galaxies and extended Ly$\alpha$ halos tracing ionised gas {\em  around} them (likely to be connected to the circumgalactic medium; CGM e.g. \citealt{Drake2019}, \citealt{Farina2019}).

\begin{table*}
	\centering
	\caption{Quasars at $z\sim6$ observed with both \ALMA\ and \MUSE. We list here: object names (column 1), coordinates (columns 2 \& 3), systemic redshift from \CII\ (z$_{\rm{[CII]}}$; column 4), three luminosities L$_{\rm{[CII]}}$ (column 5) and L$_{\rm{FIR}}$ (column 6) from \cite{Venemans20}, L$_{\rm{Ly\alpha}}$ (column 7) from \cite{Farina2019}. In the final three columns we list the diameter $d$ of the \CII, FIR and Ly$\alpha$ halo components reported in \cite{Venemans20} and \cite{Farina2019}. d$_{\rm{[CII]}}$ (column 8) and d$_{\rm{FIR}}$ (column 9) are the geometric averages of a 2D Gaussian fit, and d$_{\rm{Ly\alpha}}$ (column 10) is the diameter at which the surface-brightness-dimming-corrected light profile drops below 3 \texttimes $10^{-18}$ \SBunits\}}
	\label{Tab:1}
	\begin{tabular}{lrrrccrrrr} 
		\hline \hline
{\bf{Object}} &	{\bf{RA}} &	{\bf{DEC}} & {\bf{z}}  & {\bf{L$_{{\rm{[CII]}}}$}} & {\bf{L$_{{\rm{FIR}}}$}} & {\bf{L$_{{\rm{Ly\alpha}}}$}} & {\bf{d$_{{\rm{[CII]}}}$}}  & {\bf{d$_{{\rm{FIR}}}$}}  & {\bf{d$_{{\rm{Ly\alpha}}}$}}  \\

 	& (ICRS) & (ICRS) &  & 10$^8 L\odot$ & 10$^{11} L\odot$& erg s$^{-1}$& pkpc & pkpc & pkpc \\
\hline
{\bf{J0129-0035}} & 022.4938 & -0.5944 & 5.7788 & 1.92$\pm$0.07 & 4.76$\pm$0.11 & $<$1e+42 &  1.74 $\pm$ 0.17 &  1.10 $\pm$ 0.06 & - \\ 
 {\bf{J1044-0125}} & 161.1377 & -1.4172 & 5.7846 & 1.64$\pm$0.21 & 5.48$\pm$0.22 & $<$1e+42 &  1.88 $\pm$ 0.58 &  0.94 $\pm$ 0.12 & - \\ 
 {\bf{P007+04}} & 007.0274 & 04.9571 & 6.0015 & 1.58$\pm$0.09 & 4.52$\pm$0.11 & $<$2e+42 &    1.30 $\pm$ 0.23 &  0.65 $\pm$ 0.11 & - \\ 
 {\bf{P009-10}} & 009.7355 & -10.4317 & 6.0040 & 9.07$\pm$0.66 & 7.13$\pm$0.69 & 9e+42 &   4.04 $\pm$0.57 &  3.03 $\pm$0.46 & 20.40 $\pm$2.15 \\ 
 {\bf{J2054-0005}} & 313.5271 & -0.0874 & 6.0389 & 3.08$\pm$0.14 & 6.20$\pm$0.19 & $<$1e+42 &  1.37 $\pm$ 0.11 &  0.79 $\pm$ 0.06 & - \\ 
 {\bf{J2100-1715}} & 315.2279 & -17.2561 & 6.0807 & 1.31$\pm$0.14 & 1.12$\pm$0.16 & $<$1e+42 &  1.67 $\pm$ 0.45 &  1.30 $\pm$ 0.00 & - \\
 {\bf{J2318-3029}} & 349.6379 & -30.4927 & 6.1456 & 2.22$\pm$0.12 & 6.29$\pm$0.14 & $<$1e+42 & 1.24 $\pm$ 0.11 &  0.75 $\pm$ 0.06 & - \\
 {\bf{P359-06}} & 359.1352 & -6.3831 & 6.1719 & 2.62$\pm$0.13 & 1.60$\pm$0.15 & 3.3e+43 &  2.10 $\pm$0.22 &  1.88 $\pm$0.28 & 20.09 $\pm$2.12 \\ 
 {\bf{P065-26}} & 065.4085 & -26.9544 & 6.1871 & 1.71$\pm$0.17 & 2.80$\pm$0.23 & 6.6e+43 & 4.33 $\pm$0.95 &  1.10 $\pm$0.39 & 20.07 $\pm$2.11 \\
 {\bf{P308-21}} & 308.0416 & -21.2340 & 6.2355 & 3.37$\pm$0.19 & 2.45$\pm$0.16 & 8.8e+43 & 3.14 $\pm$0.56 &  3.20 $\pm$0.56 & 30.64 $\pm$3.23 \\
 {\bf{J0100+2802}} & 015.0543 & 28.0405 & 6.3269 & 3.76$\pm$0.17 & 2.92$\pm$0.18 & $<$5e+42 & 2.41 $\pm$ 0.55 &  1.84 $\pm$ 0.44 & - \\
 {\bf{J025-33}} & 025.6822 & -33.4627 & 6.3373 & 5.65$\pm$0.22 & 5.31$\pm$0.24 & $<$3e+42 & 2.06 $\pm$ 0.17 &  1.28 $\pm$ 0.11 & - \\ 
 {\bf{P183+05}} & 183.1124 & 05.0926 & 6.4386 & 7.15$\pm$0.32 & 10.53$\pm$0.36 & $<$6e+42 & 3.29 $\pm$ 0.27 &  2.14 $\pm$ 0.11 & - \\
 {\bf{J2318-3113}} & 349.5765 & -31.2296 & 6.4429 & 1.59$\pm$0.14 & 0.79$\pm$0.17 & $<$2e+42 &  3.32 $\pm$ 0.71 &  2.57 $\pm$ 1.65 & - \\ 
 {\bf{P036+03}} & 036.5078 & 03.0498 & 6.5405 & 3.38$\pm$0.09 & 5.77$\pm$0.12 & 3.8e+43 & 1.96 $\pm$0.11 &  0.89 $\pm$0.05 & 19.44 $\pm$2.05 \\ 
 {\bf{P231-20}} & 231.6577 & -20.8336 & 6.5869 & 3.53$\pm$0.30 & 9.99$\pm$0.34 & 1.1e+44 &  1.02 $\pm$0.11 &  0.55 $\pm$0.05 & 29.69 $\pm$3.13 \\ 
 {\bf{P323+12}} & 323.1382 & 12.2986 & 6.5872 & 1.45$\pm$0.19 & 0.53$\pm$0.27 & 2.01e+44 &  2.04 $\pm$0.54 &  0.59 $\pm$0.00 & 45.52 $\pm$4.79 \\ 
 {\bf{J0305-3150}} & 046.3205 & -31.8488 & 6.6139 & 5.90$\pm$0.36 & 12.30$\pm$0.44 & 8e+42 & 2.70 $\pm$0.22 &  1.60 $\pm$0.11 & 12.60 $\pm$1.33 \\ 
 
	\hline
	\end{tabular}
\end{table*}

In this work we examine $z\sim6$ QSOs observed with high-resolution \ALMA\ configurations to compare and contrast the bright, central emission from the host galaxies (the core-like component reported in \citealt{Novak20} and examined in \citealt{Venemans20}) to observations of the same QSOs with \MUSE\ as part of the \requiem\ survey \citep{Farina2019} which have reported the presence/absence of Ly$\alpha$ halos (tracing gaseous reservoirs) down to $5\sigma$ surface-brightness limits of $0.1 - 1.1$ $\times$ $10^{-17}$ \SBunits\ over a 1 arcsec$^2$ aperture. This paper proceeds as follows: in Section \ref{sect:obs} we describe the observations and datasets compiled for this work; in Section \ref{sect:results} we present \CII\ and Ly$\alpha$ channel maps and moment maps of our targets. In Section \ref{sect:disc} we discuss our findings and comment on our targets individually. Finally, in Section \ref{sect:concl} we summarise our findings. 

Throughout this work we assume a $\Lambda$CDM cosmology with $\Omega_m = 0.3$, $\Omega_{\Lambda} = 0.7$ and H$_0=70$ km s$^{-1}$ Mpc$^{-1}$. In this cosmology, \mbox{1\,\arcsec\ = \pkpcatsix\ pkpc} at $z \approx 6$. All velocities, wavelengths and frequencies refer to vacuum values.

\section{Observations and Sample Selection} \label{sect:obs}

\subsection{Sample Selection} \label{sect:sample}

We draw our sample from a total of  \nosample\ sources that represent the overlap between the high--resolution \ALMA\ \CII\ imaging survey published in \cite{Venemans20} and the \MUSE\ \requiem\ survey \citep{Farina2019}. We list the full sample in Table \ref{Tab:1}, before defining a subset where a Ly$\alpha$ halo detection has been made in \cite{Drake2019} or \cite{Farina2019}. The sample with robust detections in both datasets amounts to \nosubsample\ objects.

\subsection{MUSE Data} \label{sect:MUSE}
\MUSE\ data cubes are taken from the \requiem\ survey \citep{Farina2019}, in order to homogenise the reduction process across all objects. \cite{Farina2019} define the Ly$\alpha$ halo as a 3D structure of connected voxels of significance $>$\,\sigmahalo\,$\sigma$ after smoothing the PSF--subtracted datacube in the spatial and spectral directions with kernels of \kernelEma. The study also provide moment maps for the Ly$\alpha$ halos, with a velocity zero point that refers to the flux--weighted centroid of the voxels included in the mask. For the purpose of this work, we require moment maps centred on systemic velocity, and so we utilise additional data products released with \requiem\ to construct our own maps as described in Section \ref{sect:moment maps}.

\subsection{ALMA Data} \label{sect:ALMA}

\ALMA\ data cubes of the continuum--subtracted [CII] emission line are taken from \cite{Decarli2018}, \cite{Venemans2019} and \cite{Venemans20}. The data have been re--imaged with natural weighting to maximise sensitivity to any extended emission, and sampled with a velocity width of 30MHz per channel to match the velocity width of a single layer of the (slightly oversampled) \MUSE\ datacube ($\sim 40$ \kms). All data cubes were observed in the local standard of rest (LSRK) velocity frame.

\section{Results} \label{sect:results}

\subsection{Moment Maps} \label{sect:moment maps}

As we require the velocity fields (moment 1) to describe ionised gas motion relative to the systemic velocity of the host galaxy (defined as the \CII\ redshift), we utilise two data-products released with the \requiem\ survey; the PSF-subtracted datacubes, and the 3D mask cubes to reconstruct the 3D Ly$\alpha$ halo as defined in \cite{Farina2019}. With these cubes we then re-compute moment maps relative to systemic velocity, using the method applied in \cite{Drake20}. 

In Figure \ref{Fig: mom ex} we show an example of the Ly$\alpha$ and \CII\ moment maps, for \pmerg. 
 Maps for the full sample can be found in the figure set available in the online journal. The three columns of panels present the zeroth moment (total flux), the first moment (velocity field) and the second moment (a measure of the velocity dispersion), respectively. The upper row of panels focuses on the Ly$\alpha$ halo in cutouts of 10\arcsec\ a side, while the lower panels present \CII\ emission originating in the quasar host galaxy in panels which zoom--in to 4\arcsec\ a side.

An immediate take-away from this comparison is the relative sizes of the extended halo gas traced by Ly$\alpha$, and the more compact \CII\ emission which traces the host galaxy. Typically, the extent of the  \CII\ emission is encompassed within the region of $\sim$ 1\arcsec diameter, which corresponds to the region of the Ly$\alpha$ emission that is subject to complex residuals due to the PSF--subtraction (here, this region is masked on the Ly$\alpha$ images). We will return to size comparison in Section \ref{sect:sizes} where measurements for the full sample will be considered. 

The central columns of Figure \ref{Fig: mom ex} depicting the velocity fields of the emission demonstrate a general feature of the data: There is no obvious coherence between the velocity field of the host galaxy, and that of the extended ionised gas for any of the \nosubsample\ objects. We will discuss this further in Section \ref{sect: kinematics}. We also note that a comparison of the general velocity shifts is presented in section 5.1.1. and figure 5 of \cite{Farina2019}.

\subsection{Channel Maps} \label{sect:channel maps}
In addition to the moment analysis of gas kinematics for each object, we include in the Appendix a series of channel maps comparing Ly$\alpha$ and \CII\ emission for each QSO. In every case, the Ly$\alpha$ extends over a much greater velocity range than the more compact \CII\ emission, and the maps emphasise again the extent of the gaseous ionised halo gas compared to the cold gas in the host galaxy. 

\begin{figure*}
    \centering
\figurenum{2}
	\includegraphics[width=0.99\textwidth]{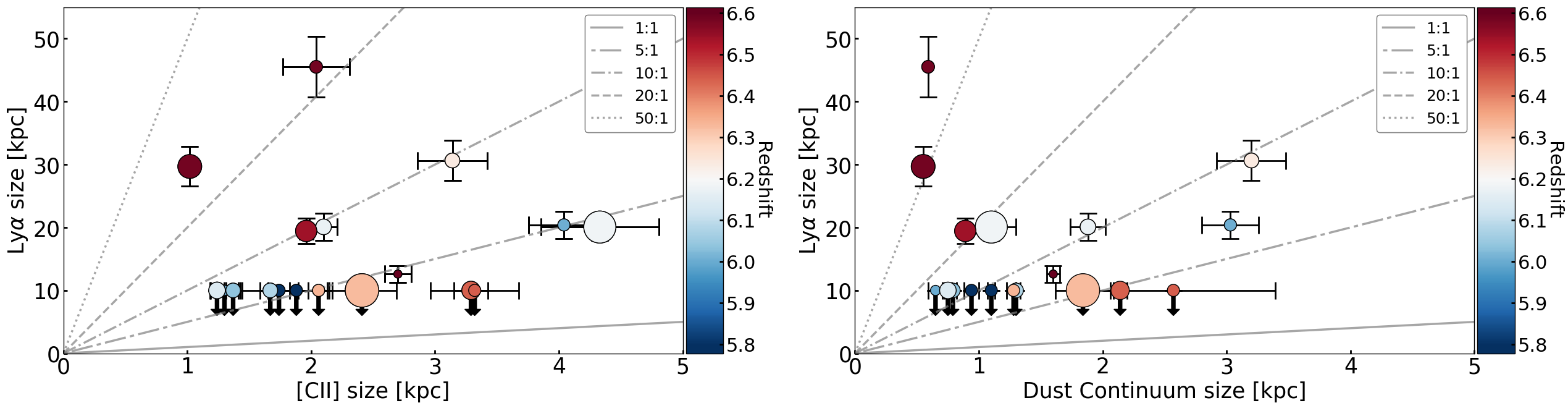}
	
	\caption{Size comparison between host galaxy in \CII\ (left panel) and FIR dust continuum (right panel) and the Ly$\alpha$ halo. \CII\ and FIR sizes are adapted from \cite{Venemans20}, Ly$\alpha$ sizes are taken from \cite{Farina2019} and represent the diameter of each halo. The size of each symbol represents its black hole mass, values taken from \cite{Schindler2021}. The diagonal grey lines represent lines of constant size ratio between the Ly$\alpha$ halo and the host galaxy. Downward arrows represent objects undetected in Ly$\alpha$ for which we assume a maximum size of 10 pkpc.}
    \label{Fig: size comp}
\end{figure*} 
\begin{figure*}
    \centering
	\figurenum{3}
	\includegraphics[width=0.99\textwidth]{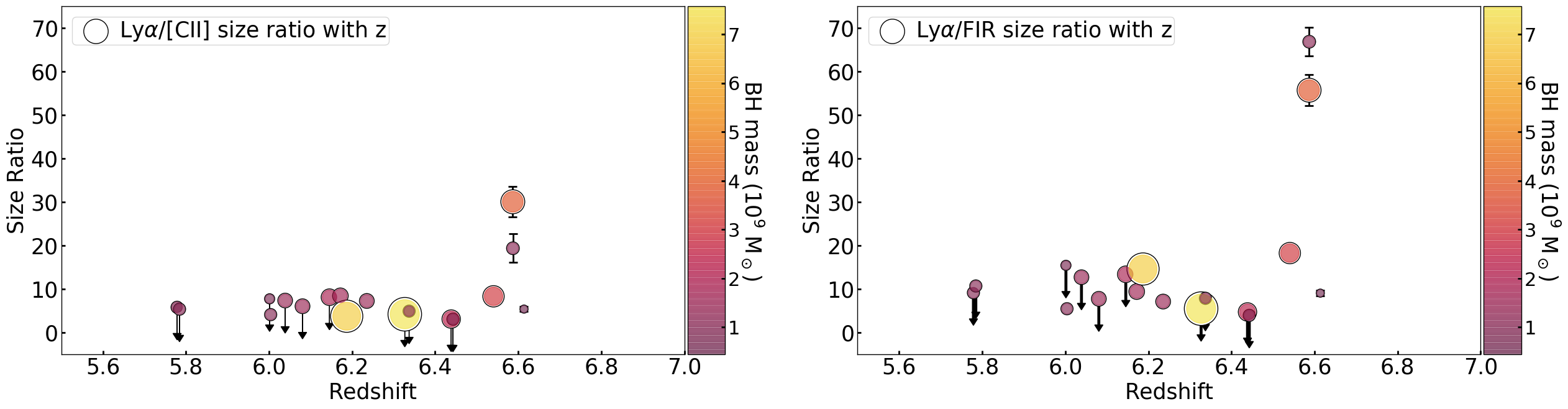}

	\caption{Size ratio between Ly$\alpha$ halo and \CII (left panel) / FIR dust continuum (right panel) as a function of redshift. Symbol sizes and colours encode the black hole mass, downward arrows represent objects undetected in Ly$\alpha$ for which we assume a maximum size of 10 pkpc.}
    \label{Fig: size ratio}
\end{figure*}

\section{Discussion} \label{sect:disc}

\subsection{Size Comparison} \label{sect:sizes}

In Figure \ref{Fig: size comp} we compare the sizes of the \CII\ and FIR continuum emission to that of the Ly$\alpha$ halo. The adopted size measurements are listed in Table \ref{Tab:1}. For the Ly$\alpha$ size we take the surface--brightness--dimming corrected size from \cite{Farina2019} (their method 4) for consistency of method across all objects and redshifts. For the \CII\ and FIR sizes we take values from \cite{Venemans20}, where sizes from the {\sc{casa}} task {\sc{imfit}} are provided using a 2D fit. In this work we take the geometric average of these two measurements as the effective \CII\ or FIR extent.

The two panels show on the x-axes the \CII\ size (left--hand side) and FIR dust continuum (right--hand size) against the Ly$\alpha$ extend on the y--axes. As was evident from Figure \ref{Fig: mom ex} the Ly$\alpha$ halos are significantly more extended than the \CII\ (and also dust continuum) emission. This is because the \CII\ emission primarily traces photo--dominated regions within the host galaxy, meanwhile, the extended Ly$\alpha$ emission traces the extended ionized gas surrounding the central host galaxy. 

Ly$\alpha$ halo sizes fall anywhere between $\sim 2$ and $\sim 30$ times larger than the \CII\ sizes. \poog, at $z=6.00$ shows the smallest Ly$\alpha$ halo, $4$ times larger than its extended \CII\ which stretches $\sim4$ pkpc. At the other end of the scale, \pchiara, at $z=6.59$ has a Ly$\alpha$ halo extending $\approx 30$ times further than its compact \CII\ emission which is measured at $1$ pkpc. In the right-hand panel we confirm that the FIR (tracing the dust continuum) is smaller than the \CII\ size (e.g. see \cite{Novak20} for a detailed analysis of \CII\ and FIR dust continuum sizes). The Ly$\alpha$ halos range between $\sim 3$ and $> 50$ times larger than the dust continuum. Much as for the \CII\ comparison, \poog\ exhibits the lowest size ratio of $d_{{\rm{Ly\alpha}}}/d_{{\rm{FIR}}} = 5$, and again, \pchiara\ exhibits the largest size ratio of $d_{{\rm{Ly\alpha}}}/d_{{\rm{FIR}}} = 67$. Interestingly, in the FIR,  \pthreetothree, moves up the ranking, and exhibits the second largest size ratio $d_{{\rm{Ly\alpha}}}/d_{{\rm{FIR}}} = 61$. Additional information is encoded in the figure, using QSO redshift to colour the data points, and scaling the size of each point by the QSO's SMBH mass (taken from \citealt{Schindler2021}). No correlation is seen between M$_{{\rm{BH}}}$ and any of the size measures (Ly$\alpha$, \CII, or FIR), nor with the ratio of Ly$\alpha$ halo size to host-galaxy tracer size. A Kendall's rank correlation test (accounting for upper limits) shows that there is no evidence for a correlation between the [CII]/FIR and Ly$\alpha$ halo sizes.

In Figure \ref{Fig: size ratio} we consider the ratio of sizes between the Ly$\alpha$ halo and the two tracers of the galaxy's ISM as a function of redshift. Neither the Ly$\alpha$/\CII\ size, nor the Ly$\alpha$/FIR size shows any correlation with redshift ($\tau = 0.067$ with $p = 0.42$, and $\tau = 0.04$ with $p = 0.40$ respectively) - this implies no evolution of the size ratio across our redshift range.

\subsection{Kinematic Comparison} \label{sect: kinematics}

We now compare the overall kinematics of the ISM in the host galaxy, traced by \CII, with the larger--scale kinematics as traced by the Ly$\alpha$ line. The underlying question is whether the bulk motion between the two tracers is the same, i.e. if they trace the same gravitationally bound structure. We acknowledge that the gas kinematics traced by Ly$\alpha$ are difficult to interpret due to the complex radiative transfer of Ly$\alpha$ photons that may prevent a clear kinematic signature. 

By comparing the \CII\ velocity fields to those of the Ly$\alpha$ line we find, to first order, no evidence for coherent rotation between these two tracers. In many cases this is due to the fact that the Ly$\alpha$ velocity field does not show a clear velocity gradient. The \CII, on the other hand, does indeed show such a velocity gradient in the majority of sources \citep{Neeleman2021}. There are three cases, discussed in detail below, where we see a clear gradient in the velocity field of the Ly$\alpha$ halos, these are \wolf, \pthreesix, and \pchiara. In the notes below, we use the commonly--adopted definition of the position angle (P.A.) -- the angle measured from North, counter--clockwise, to the receding part of the emission.

\subsubsection{Notes on the Kinematics of Individual Objects} \label{sect:individual objs}

\subsubsection{\poog} 
\label{sect: poog}
Given the asymmetric extended \CII\ emission, \cite{Venemans20} speculate that \poog\ may be in the process of merging with another source.
Examining the moment maps in Figure \ref{Fig: mom ex}, \poog's elongation in \CII\  aligns with the orientation of the Ly$\alpha$ halo. In terms of kinematics, a velocity gradient is seen along the \CII\ emission, meanwhile an obvious gradient in the Ly$\alpha$ halo is absent. 

\subsubsection{\wolf} 
\label{sect: wolf}
The \CII\ emission from \wolf\ is surrounded by a Ly$\alpha$ halo.
The kinematics of \wolf\ in \CII\ show a velocity gradient approximately from east to west (P.A. of $315.7\pm2.6^{\circ}$, \cite{Neeleman2021}), meanwhile the kinematic structure of the surrounding Ly$\alpha$ halo does not appear to be aligned, with a mild gradient from north to south and additional blue--shifted emission in the southern outskirts. 
\subsubsection{\psixfive} 
\label{sect: 065}

\cite{Venemans20} note that the \CII\ morphology of \psixfive\ is extended and disturbed.
The Ly$\alpha$ halo surrounds the \CII, with a marginal extension towards the north. The kinematics show no obvious velocity gradient across the structure, but simply patches of mildly blue-- and red--shifted emission. In contrast, the entirety of the Ly$\alpha$ halo appears to be blue-shifted by a few hundred km\,s$^{-1}$.

\subsubsection{\pmerg} 
\label{sect: 308}

\pmerg\ is extended in both \CII\ and Ly$\alpha$ emission. The \CII\ emission has been extensively studied in \cite{Decarli2019} with high resolution imaging. The authors argue that the extended tails of \CII\ emission are due to the tidal stripping of a satellite galaxy. 
In the moment maps, we see that the Ly$\alpha$ halo is much larger still than the extended \CII. The peak of the Ly$\alpha$ emission is north-easterly from the peak in \CII\ which is centred on the QSO position. 
The \CII\ emission from the QSO host galaxy shows a velocity gradient in the south-north direction. The Ly$\alpha$ halo meanwhile is red-shifted across almost the entirety of the structure, except for the very outskirts of the halo in the east. 

\subsubsection{\pthreesix} 
\label{sect: 036}

\pthreesix\ shows a relatively compact \CII\ morphology, and a Ly$\alpha$ halo extending towards the north. The \CII\ emission shows a gradient from north to south (P.A. $189.9^{\circ}$\,$^{+1.8}_{-2.0}$, \cite{Neeleman2021}), meanwhile the Ly$\alpha$ halo  shows a gradient that is approximately perpendicular to this (from east to west, with an approximate P.A. of \PALyathreesix). \pthreesix\ presents the most striking contrast between the halo and ISM velocity fields of the objects studied here. 

\subsubsection{\pchiara} 
\label{sect: 231}

\pchiara\ has a gas-rich companion detected in \CII, presented in \cite{Decarli2018} 9\,kpc south of the QSO, and a second, fainter, companion, presented in \cite{Neeleman2019}. The \CII\ emission that is associated with the quasar is compact, and \cite{Neeleman2021} report a position angle of $83\pm4^{\circ}$. The Ly$\alpha$ halo is present prominently towards the north of the QSO. There is a pronounced velocity gradient in the east - west direction across the Ly$\alpha$ halo, with a position angle of approximately \PALyachiara.

\subsubsection{\pthreetothree} 
\label{sect: 323}

\pthreetothree\ shows very compact and faint \CII\ emission. In contrast, its Ly$\alpha$ halo is bright, and is one of the more extended of the sample.
The \CII\ kinematics do not show a velocity gradient, and the majority of the Ly$\alpha$ halo is blue-shifted.

\subsubsection{\jema} 
\label{sect: J0305}

\jema\ was first presented in \cite{venemans2016}, and supplemented with very high resolution data in \cite{Venemans20}, resolving scales of $\sim 400$ pc. The Ly$\alpha$ halo was the first $z\sim6$ halo reported with \MUSE, presented in \cite{Farina2017}. 
Neither the \CII\ emission, nor the Ly$\alpha$ halo are particularly luminous. In terms of kinematics, the \CII\ displays a clear gradient from the south west to the north east. The Ly$\alpha$ halo within which it is embedded is red-shifted in its entirety, extending mainly to the south.

\section{Summary} \label{sect:concl}

We have presented a comparison of Ly$\alpha$ and \CII\ emission of a sample of \nosubsample\ QSOs at $z\sim6$, using \MUSE\ and \ALMA\ datacubes matched in velocity resolution. The \CII\ emission traces the extent and kinematics of the interstellar medium of the QSO host--galaxy, whereas the Ly$\alpha$ emission traces the extent and kinematics of  the ionised gaseous halos that surround the quasar hosts. 

We find that the Ly$\alpha$ halo sizes are typically $3-30$ times larger than the extent of the \CII\ that is associated with the host galaxy (and $3-60$ times larger than the host galaxy's dust continuum emission). A comparison of the kinematics has proven more difficult, as the majority of the Ly$\alpha$ halos do not show ordered motion in their velocity fields. In those three cases where a kinematic P.A. can be determined in the respective Ly$\alpha$ halos, their velocity fields are not aligned with that of the [CII] emission. In other words, we find not a single case where  the rotational signature associated with the host galaxy extends to the  Ly$\alpha$ halo. This suggests that the Ly$\alpha$ emission is not simply tracing the outskirts of a large rotating disk structure that is a simple extension of the central structure seen in \CII\ (and dust) emission. It rather suggests that the kinematics of the halo gas are decoupled from those of the interstellar medium in the host galaxies' disks.

Connecting the kinematics of \CII\ and Ly$\alpha$ remains a challenge, in particular in the presence of companions and/or asymmetries in the gas distribution. An additional caveat is that, given its highly scattering nature, the Ly$\alpha$ emission line is not an ideal tracer for gas kinematics. While we do not think this should affect our ability to detect rotational signatures at large galacto-centric radii, a confirmation of our results will have to await observations of the sources with the NIRSpec IFU on--board the James Webb Space Telescope. Ideally, a large sample of isolated sources is required, to investigate the general case of cosmological accretion using the H$\alpha$ emission line, allowing a direct and unambiguous comparison of the halo kinematics to those of the host galaxy in \CII.

\appendix
\label{App}

\section{Moment Maps}
We include here the full series of moment maps continuing as in Figure 1. 

\figurenum{1 contd}
\begin{figure*}
\label{fig: all moms contd}
\includegraphics[width=0.99\textwidth]{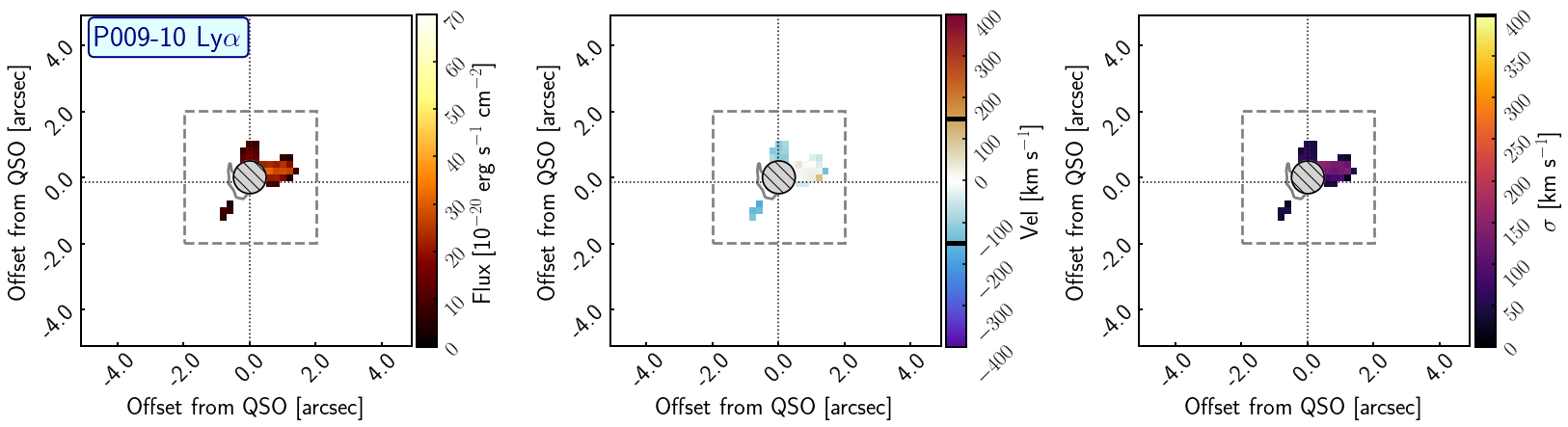}
\includegraphics[width=0.99\textwidth]{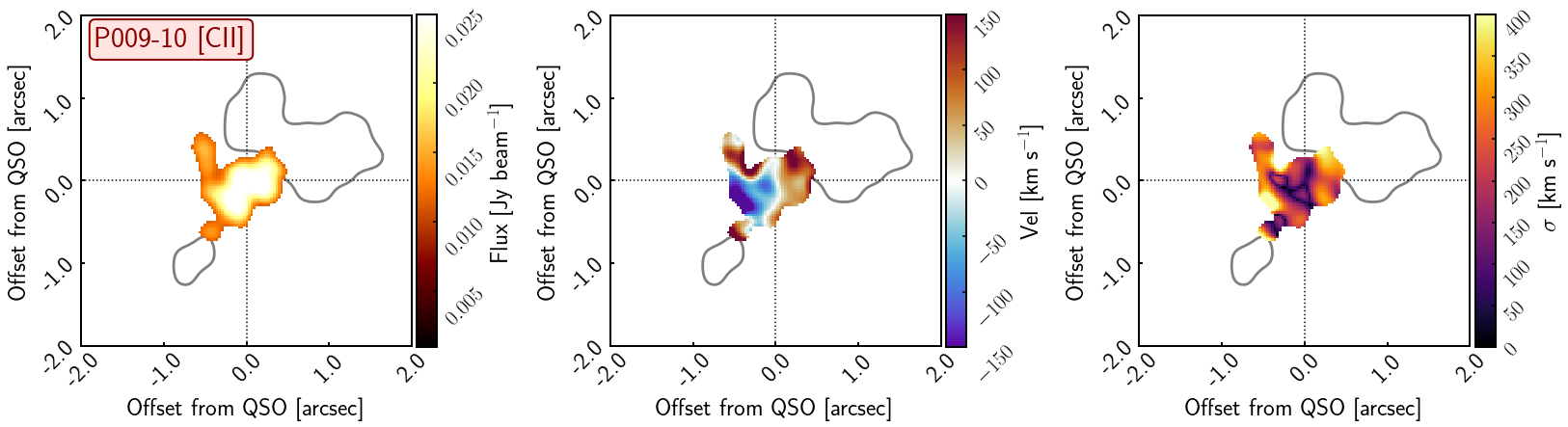}
\includegraphics[width=0.99\textwidth]{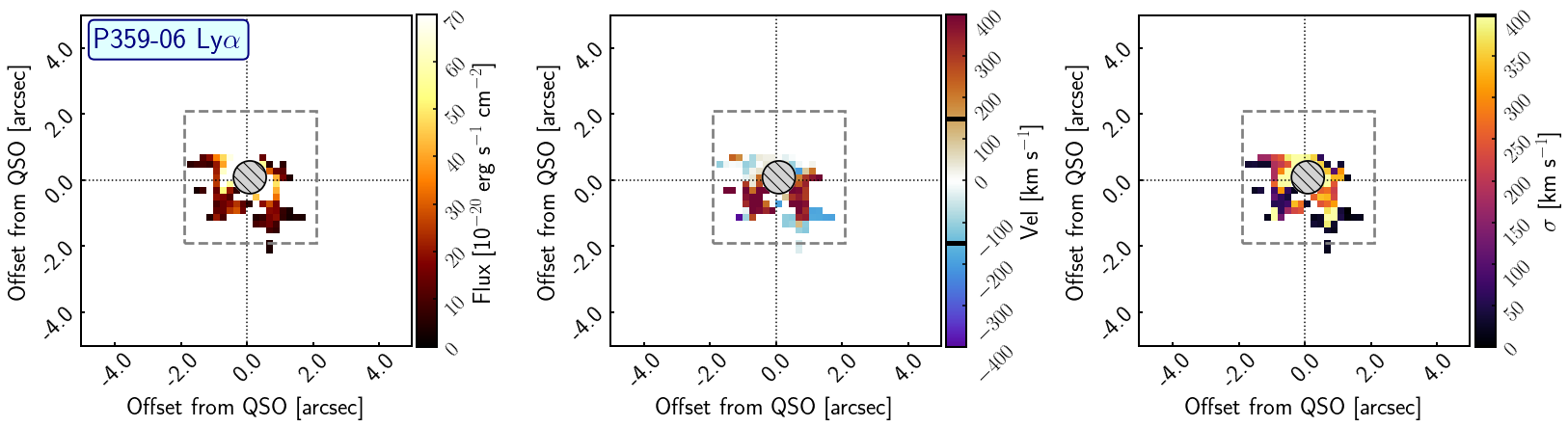}
\includegraphics[width=0.99\textwidth]{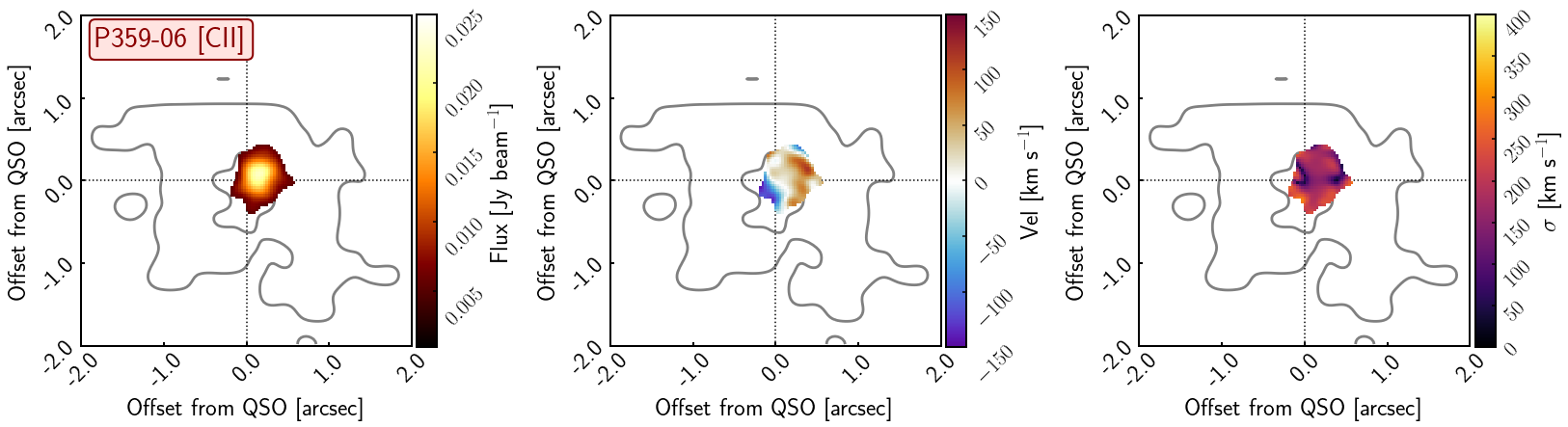}
\caption{Moment maps for the Ly$\alpha$ halo (upper panels) and \CII\ emission (lower panels) for a single QSO as in Figure 1.}
\end{figure*}

\begin{figure*}\ContinuedFloat
\includegraphics[width=0.99\textwidth]{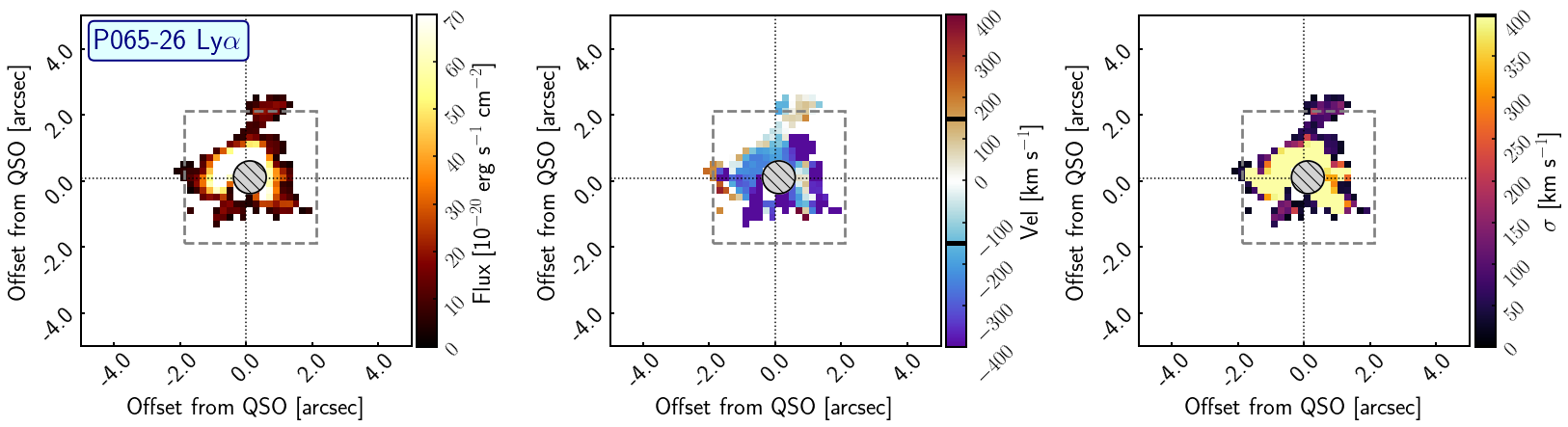}
\includegraphics[width=0.99\textwidth]{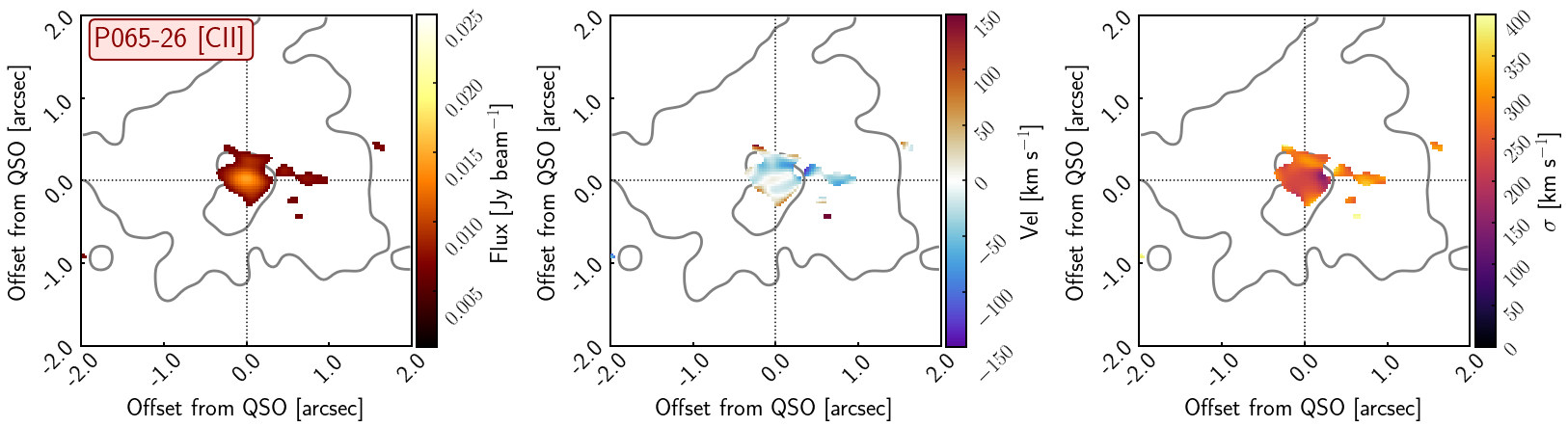}
\includegraphics[width=0.99\textwidth]{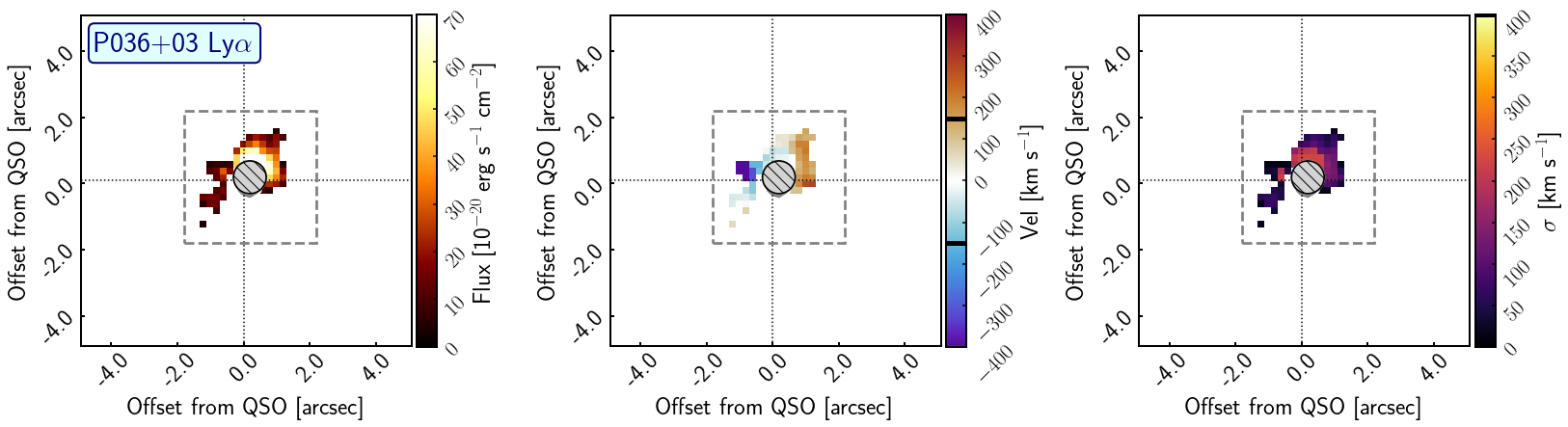}
\includegraphics[width=0.99\textwidth]{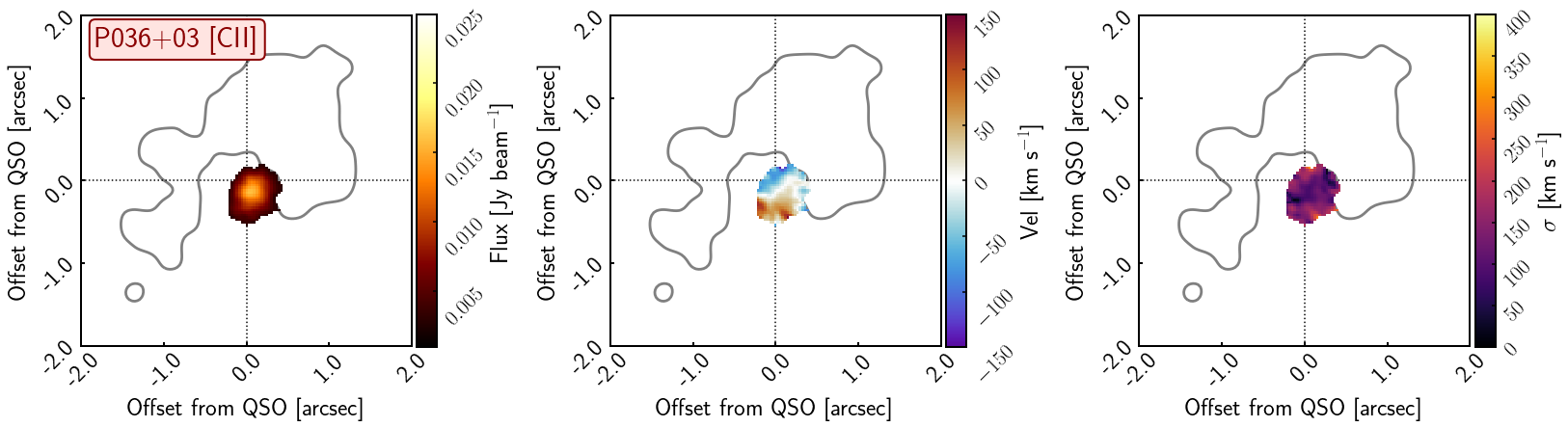}
\caption{Moment maps for the Ly$\alpha$ halo (upper panels) and \CII\ emission (lower panels) for a single QSO as in Figure 1.}
\end{figure*}

\begin{figure*}\ContinuedFloat
\includegraphics[width=0.99\textwidth]{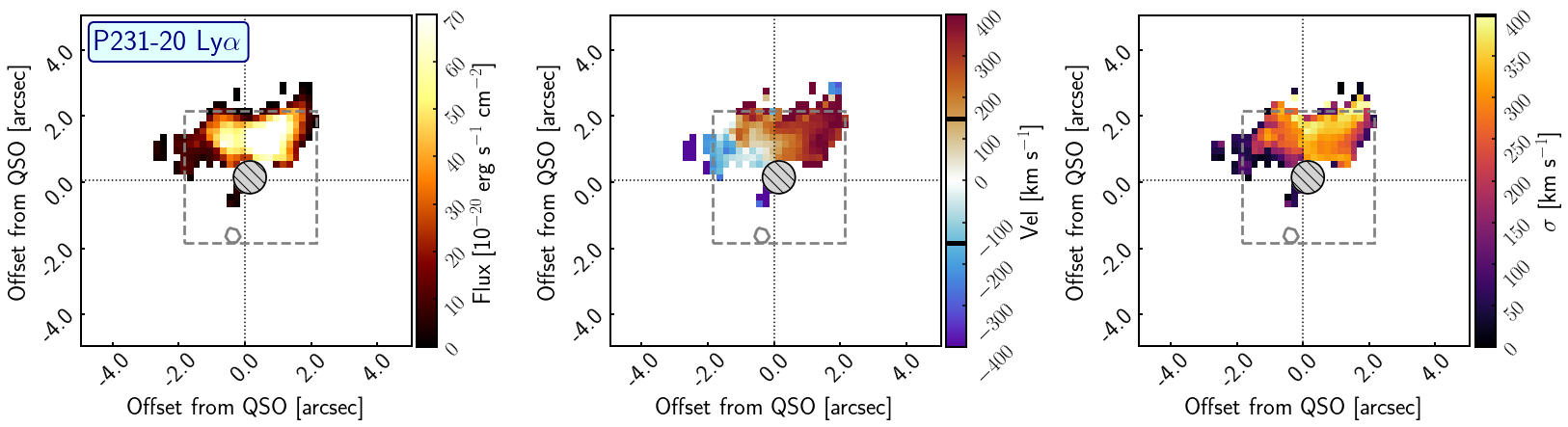}
\includegraphics[width=0.99\textwidth]{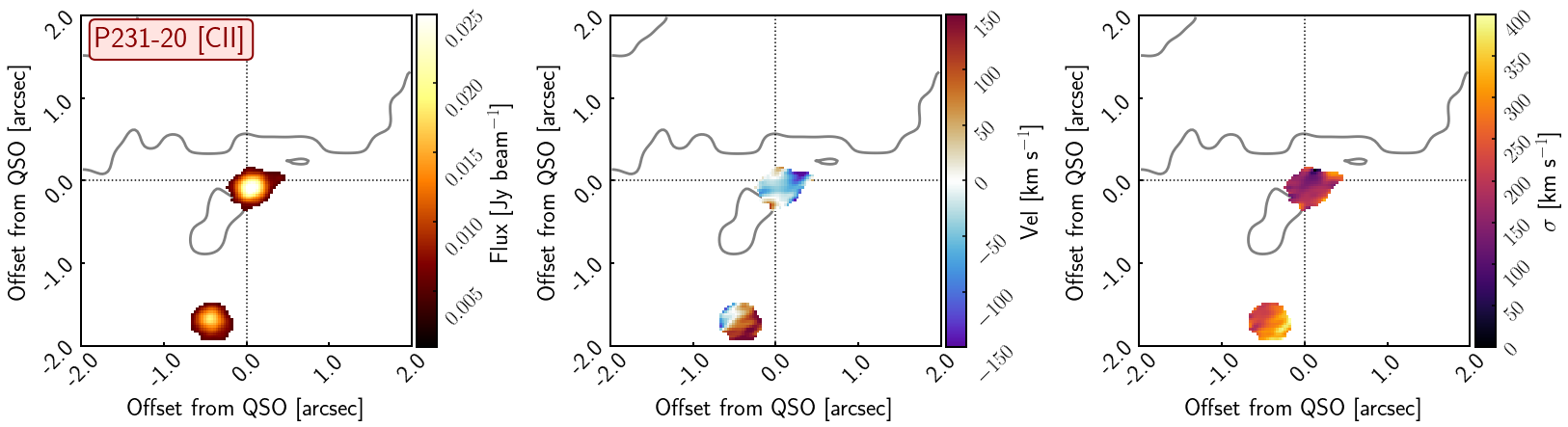}
\includegraphics[width=0.99\textwidth]{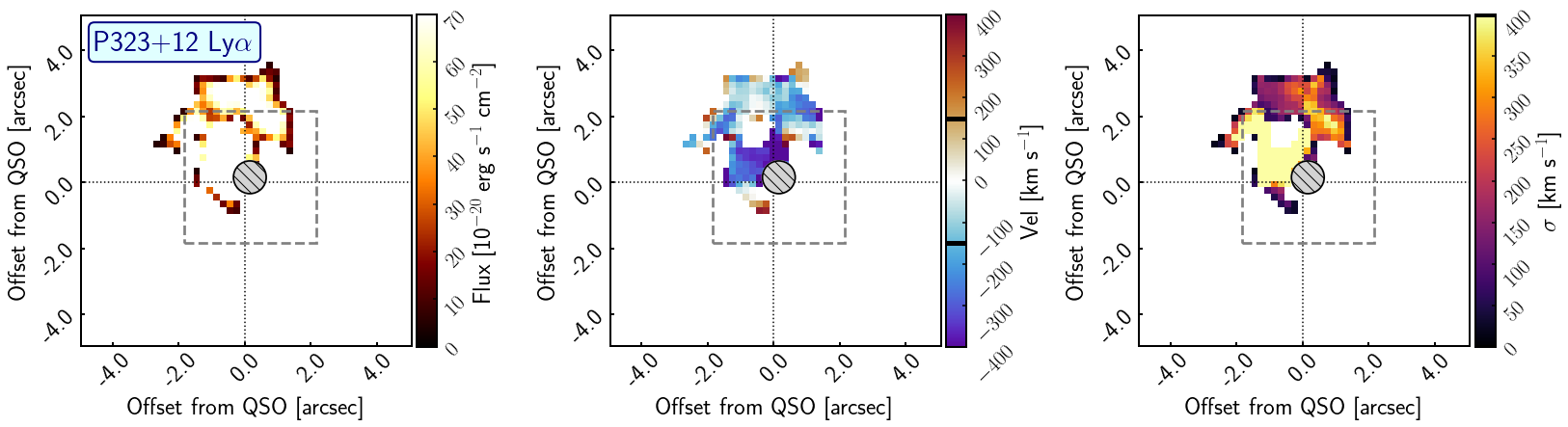}
\includegraphics[width=0.99\textwidth]{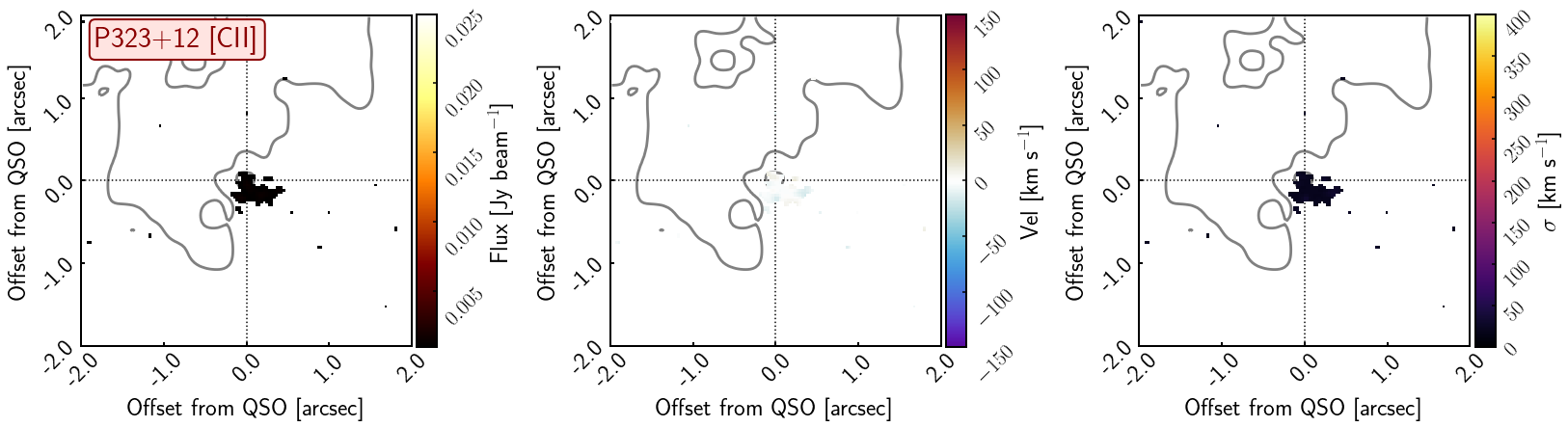}
\caption{Moment maps for the Ly$\alpha$ halo (upper panels) and \CII\ emission (lower panels) for a single QSO as in Figure 1.}
\end{figure*}

\begin{figure*}\ContinuedFloat
\includegraphics[width=0.99\textwidth]{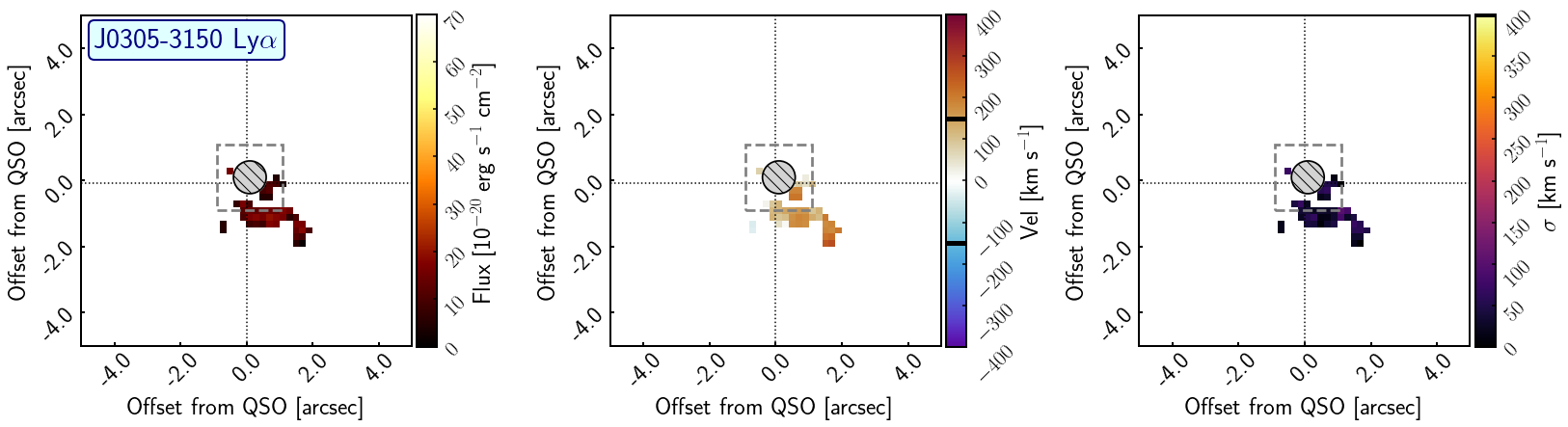}
\includegraphics[width=0.99\textwidth]{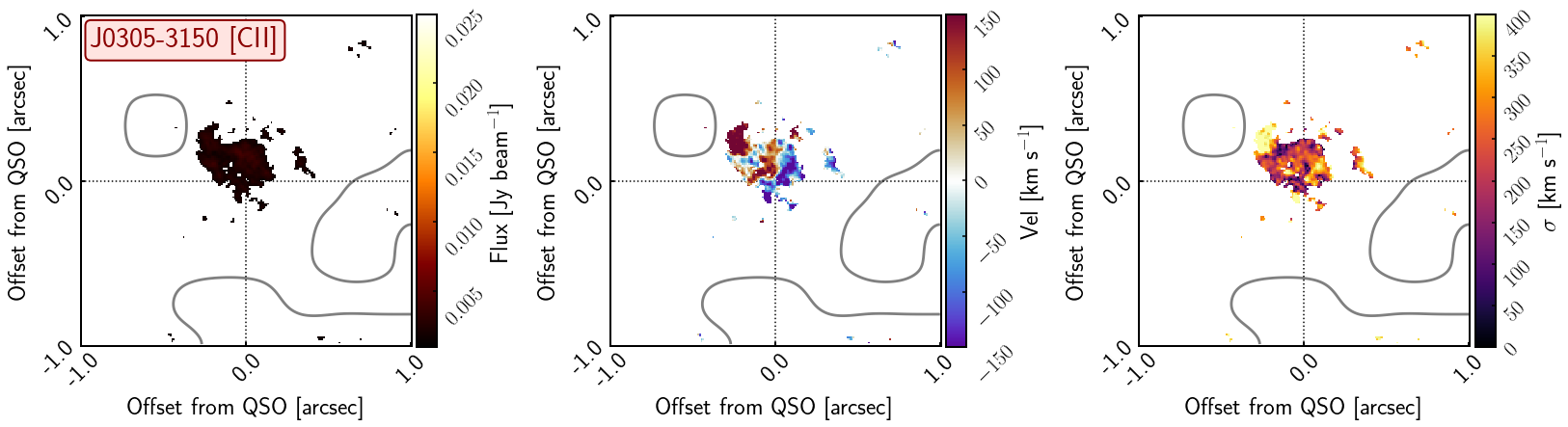}
\caption{Moment maps for the Ly$\alpha$ halo (upper panels) and \CII\ emission (lower panels) for a single QSO as in Figure 1.}
\end{figure*}

\section{Channel Maps of Ly$\alpha$ and \CII\ Emission}
We include here a series of channel maps depicting emission from \CII\ in the QSO host galaxies, and Ly$\alpha$ in the extended halos extracted at the same velocity. The [CII] emission in each panel is that arising in a single channel of 30MHz ($\sim 40$ \kms), in comparison, the Ly$\alpha$ emission from the slightly over-sampled \MUSE\ datacube is the sum across a $2$\AA\ wide window ($\sim 66$ \kms). All panels are cutouts of $10$\arcsec\ a-side, with the exception of the highest redshift object, \jema\ where the very high resolution observations from \ALMA\ dictate that we zoom in to panels of $6$\arcsec\ a-side, and remove the masking of PSF residuals in the Ly$\alpha$ image in order to see the host galaxy in \CII\ in these maps. Channel maps for the full sample can be found in the figure set available in the online journal (or in the Appendix on arXiv).\\

\figurenum{4}
\begin{figure*}
\includegraphics[width=0.99\textwidth]{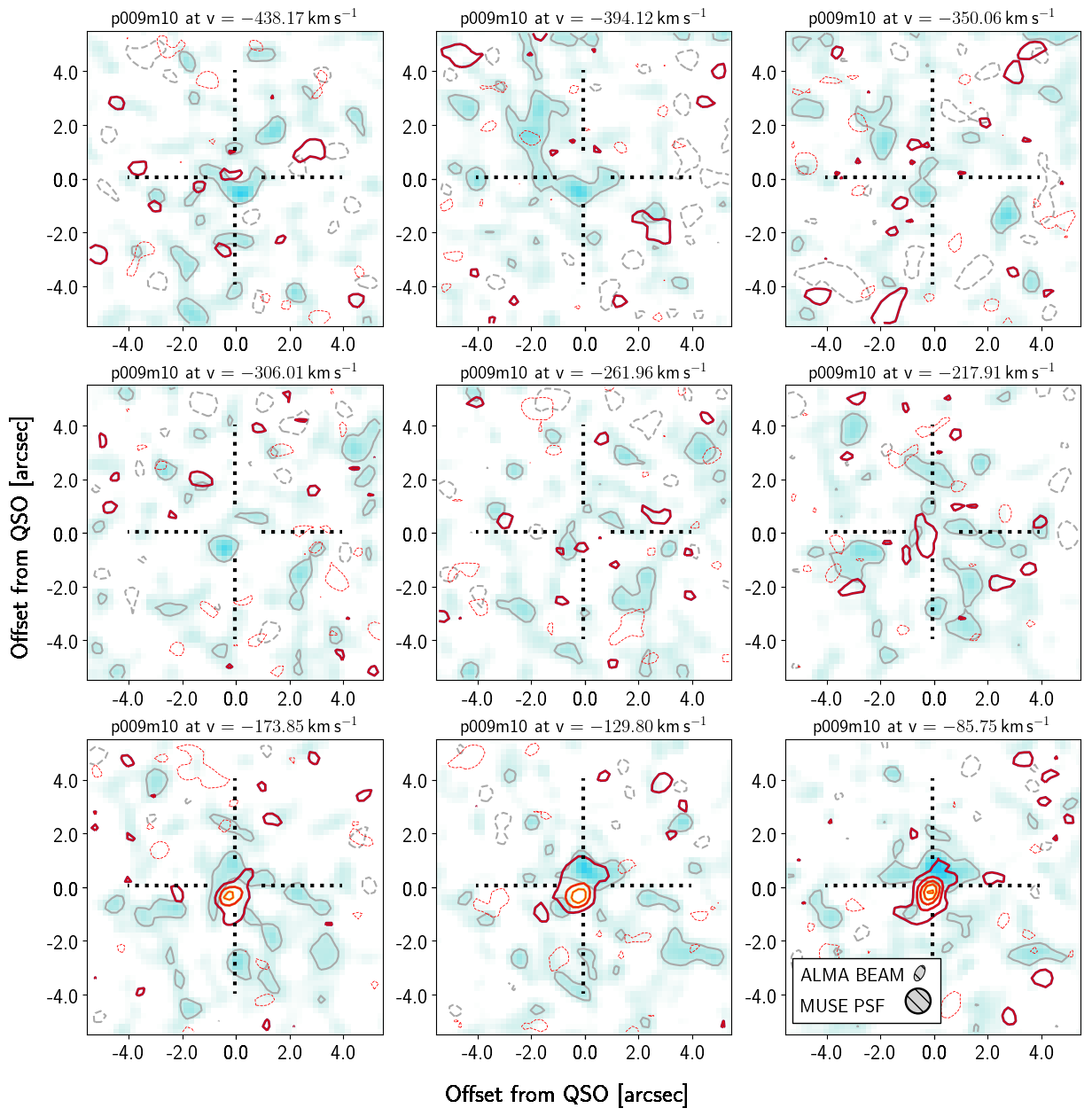}
  \caption{Channel maps displaying images of extended Ly$\alpha$ emission overlaid with linearly spaced contours depicting \CII\ emission at the same velocity. The lowest positive (solid) and negative (dashed) contour levels represent $\pm 2 \sigma$ ($\pm 4 \sigma$ for \jema) in the \CII\ channel after smoothing with a Gaussian kernel of {\sc{fwhm}} = 2.23 pixels. The \CII\ emission in each panel is that arising in a single channel of 30MHz ($\sim 40$ \kms), in comparison, the Ly$\alpha$ emission from the slightly over-sampled \MUSE\ datacube is the sum across a $2\AA$ wide window ($\sim 66$ \kms). The Ly$\alpha$ image is smoothed with a Gaussian kernel of {\sc{fwhm}} = 3.30 pixels and contoured at $\pm1.5 \sigma$ (solid, dashed lines). This demonstrates that an extended low-surface-brightness component is contiguous over multiple velocity channels. We present two pages of channel maps for each object in turn, beginning with the lowest redshift quasar, \poog. Channel maps for the full sample can be found in the figure set available in the online journal or in the Appendix on arXiv.}
\end{figure*}

\begin{figure*}\ContinuedFloat
\includegraphics[width=0.99\textwidth]{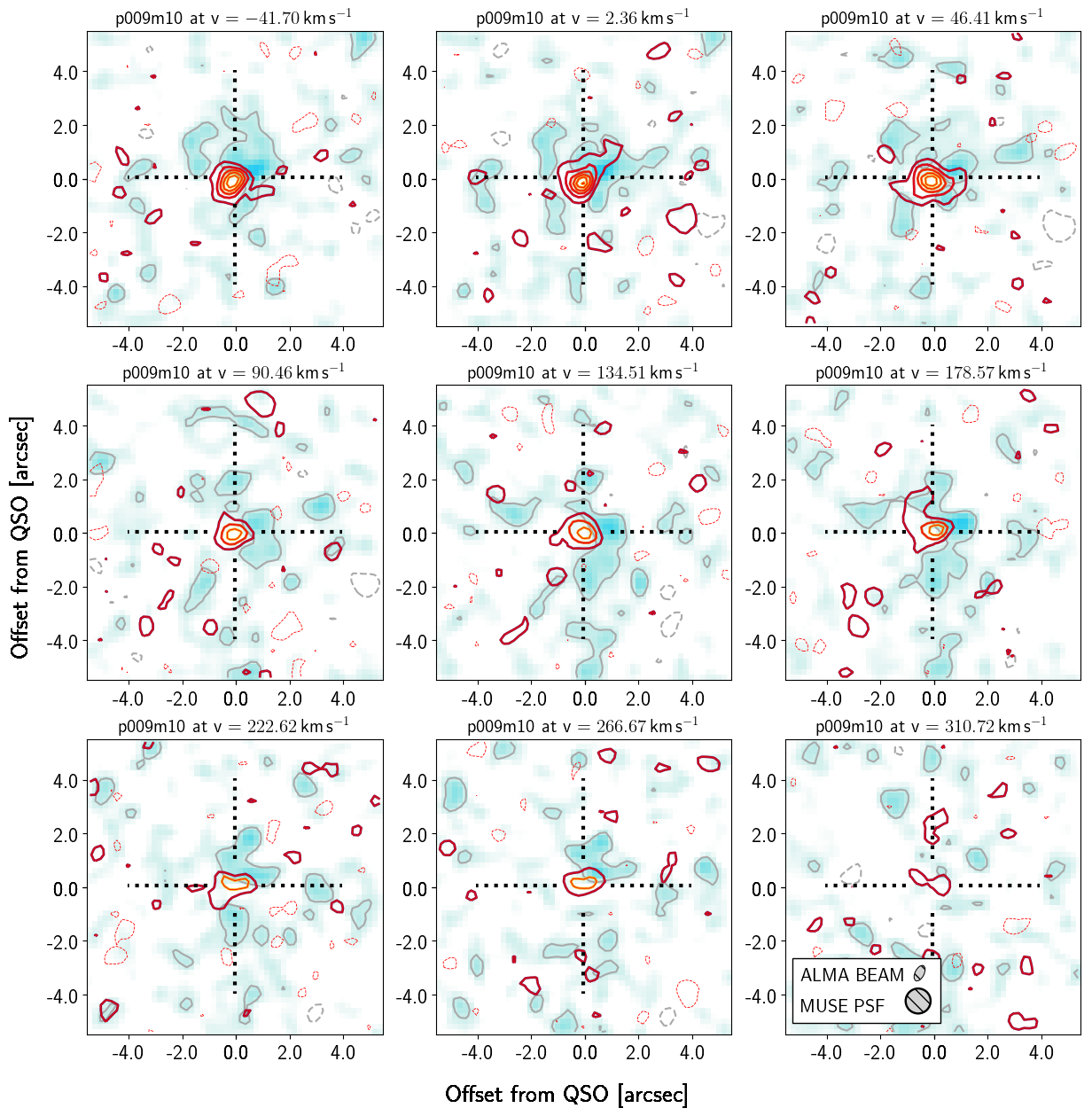}
\caption{\poog\ - continued from previous page}
\end{figure*}

\begin{figure*}\ContinuedFloat
\includegraphics[width=0.99\textwidth]{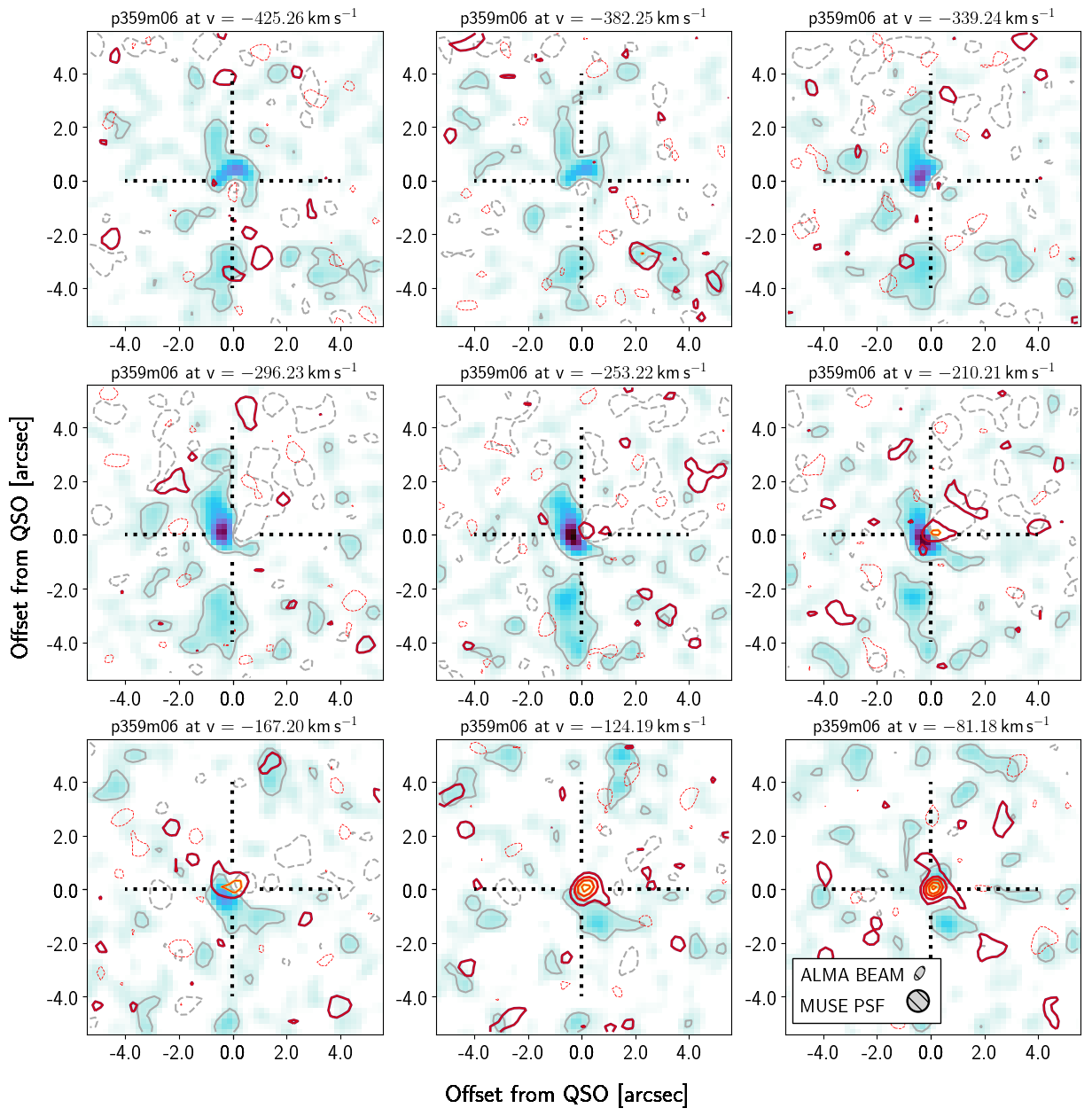}
\caption{\wolf}
\end{figure*}

\begin{figure*}\ContinuedFloat
\includegraphics[width=0.99\textwidth]{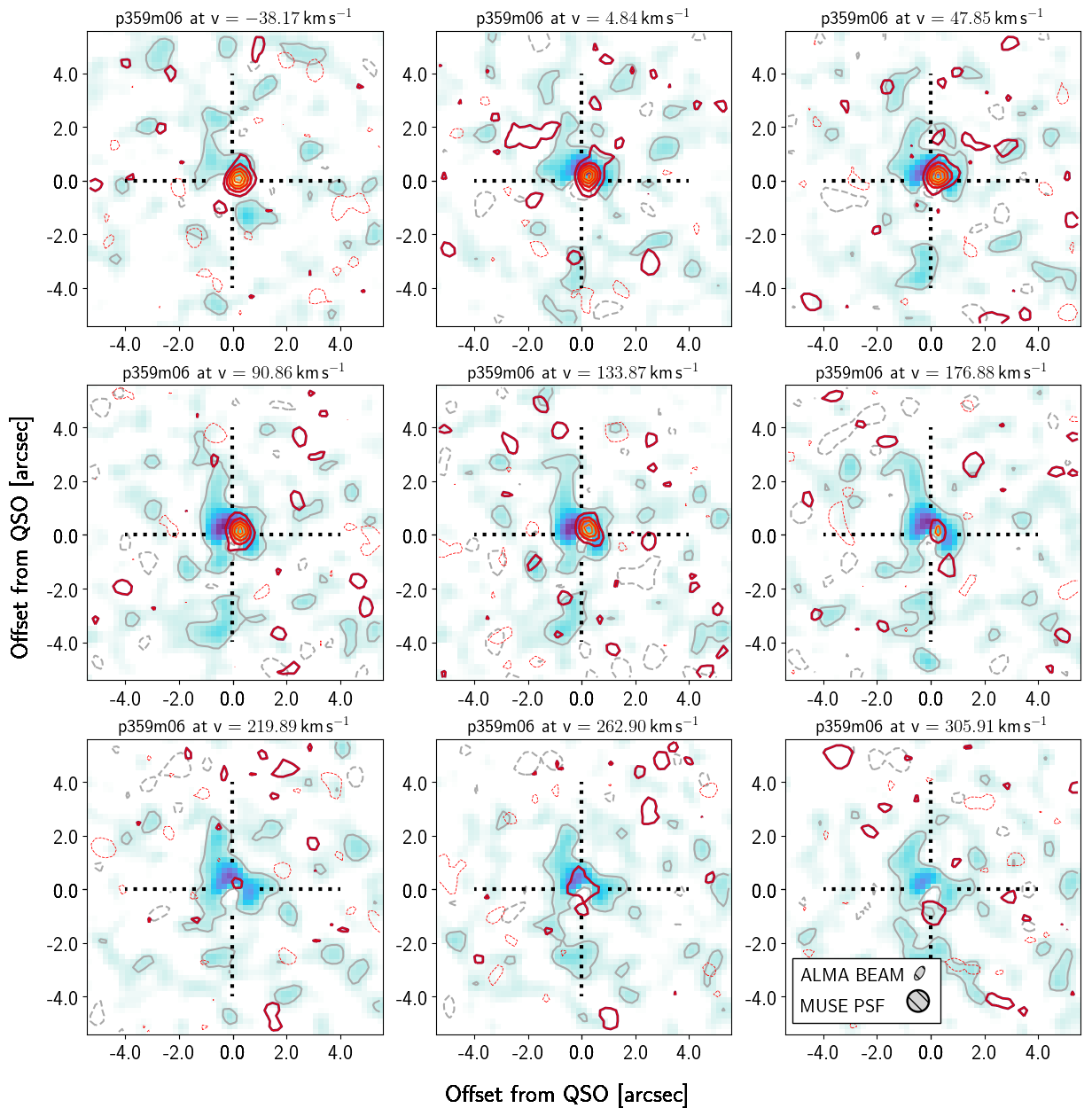}
\caption{\wolf\ - continued from previous page}
\end{figure*}

\begin{figure*}\ContinuedFloat
\includegraphics[width=0.99\textwidth]{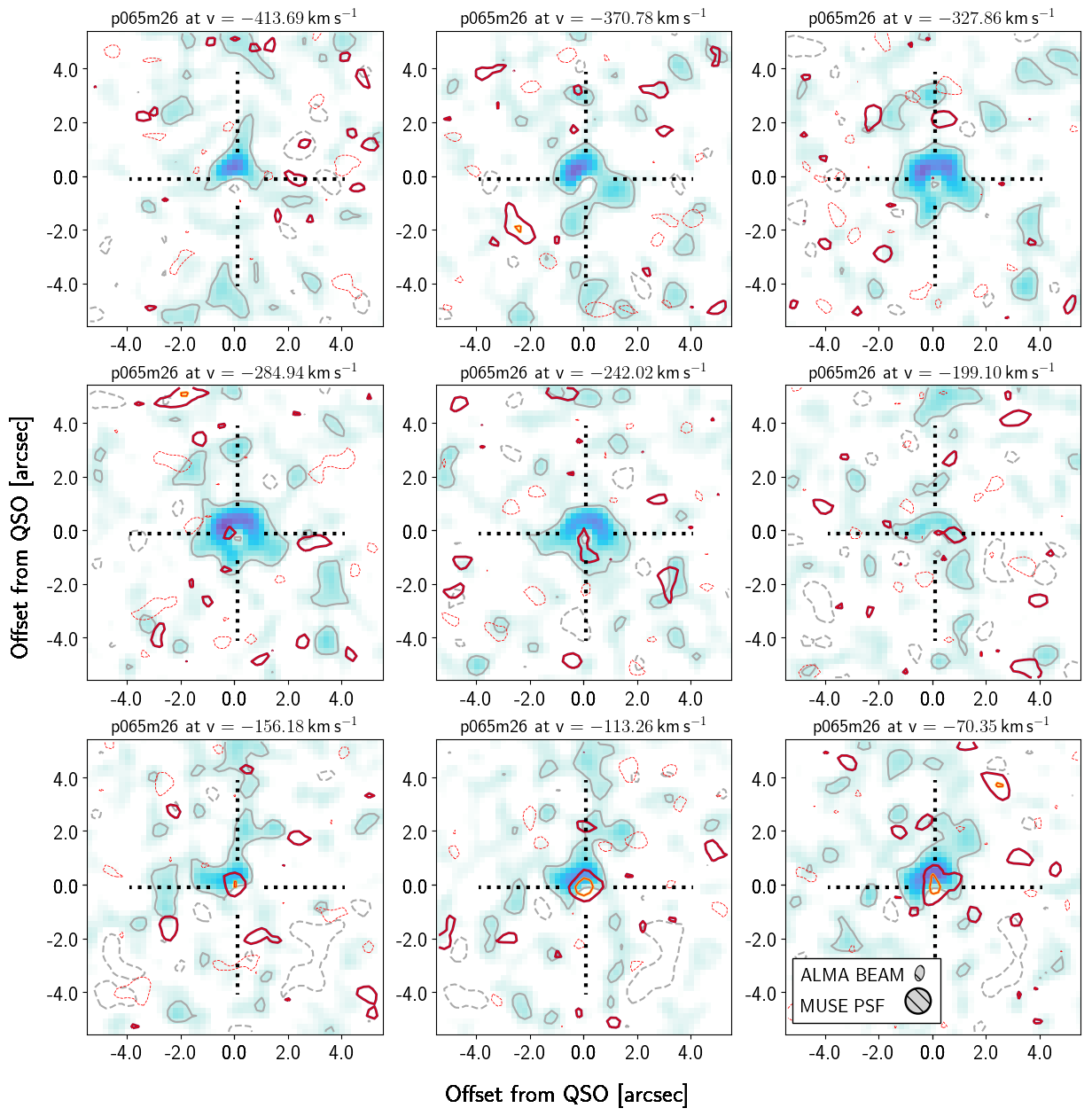}
\caption{\psixfive}
\end{figure*}

\begin{figure*}\ContinuedFloat
\includegraphics[width=0.99\textwidth]{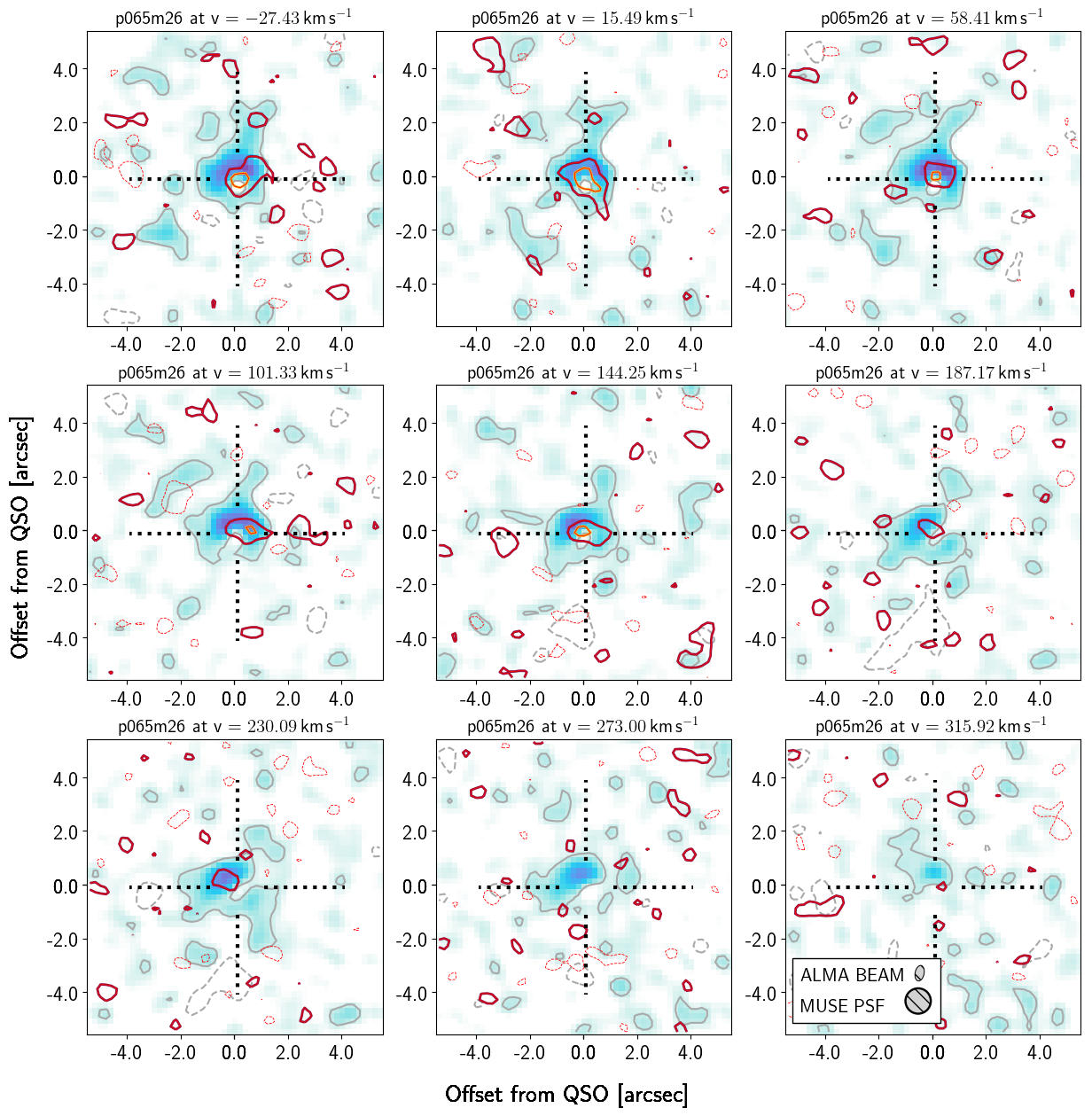}
\caption{\psixfive\ - continued from previous page}
\end{figure*}

\begin{figure*}\ContinuedFloat
\includegraphics[width=0.99\textwidth]{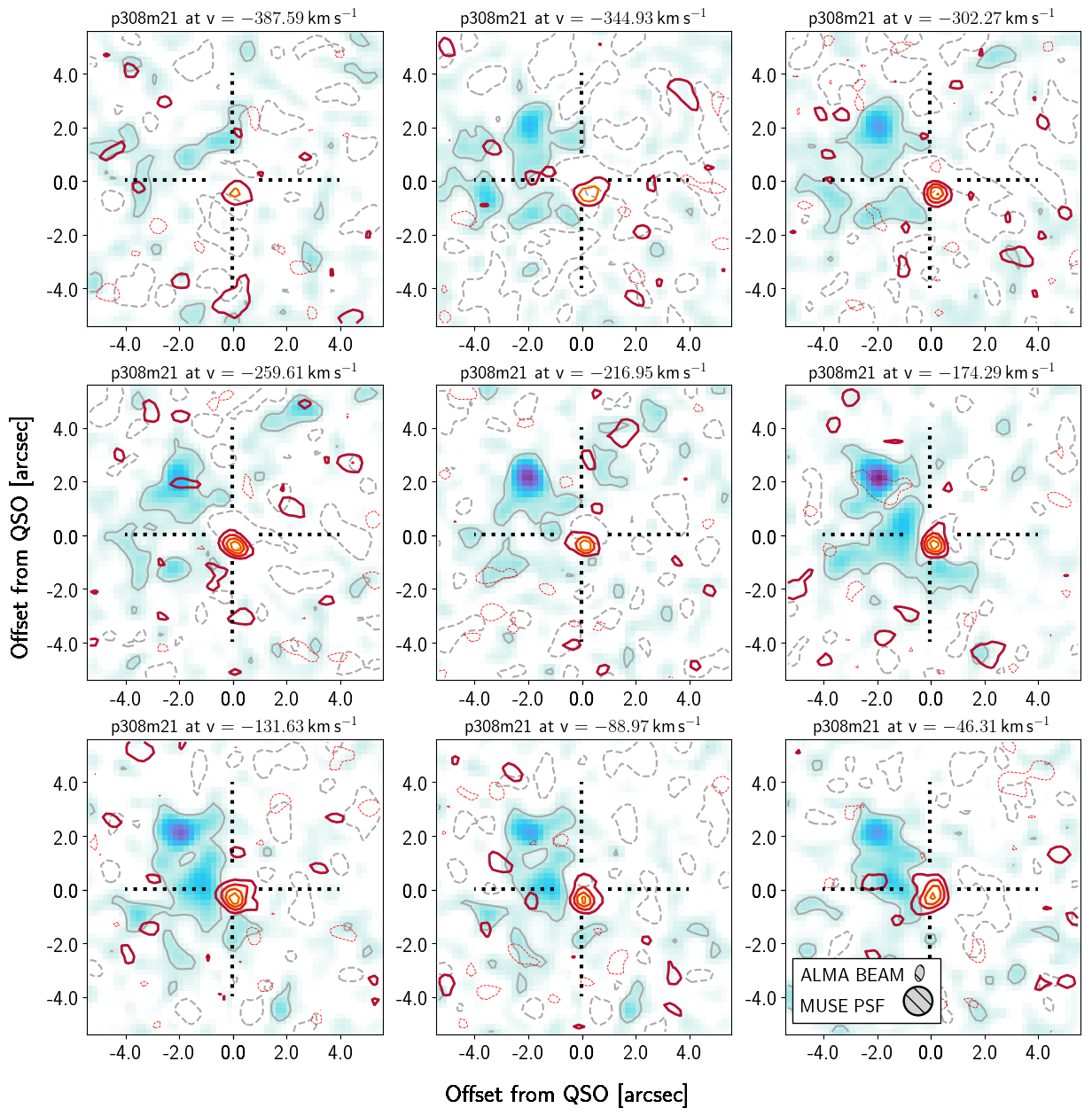}
\caption{\pmerg}
\end{figure*}

\begin{figure*}\ContinuedFloat
\includegraphics[width=0.99\textwidth]{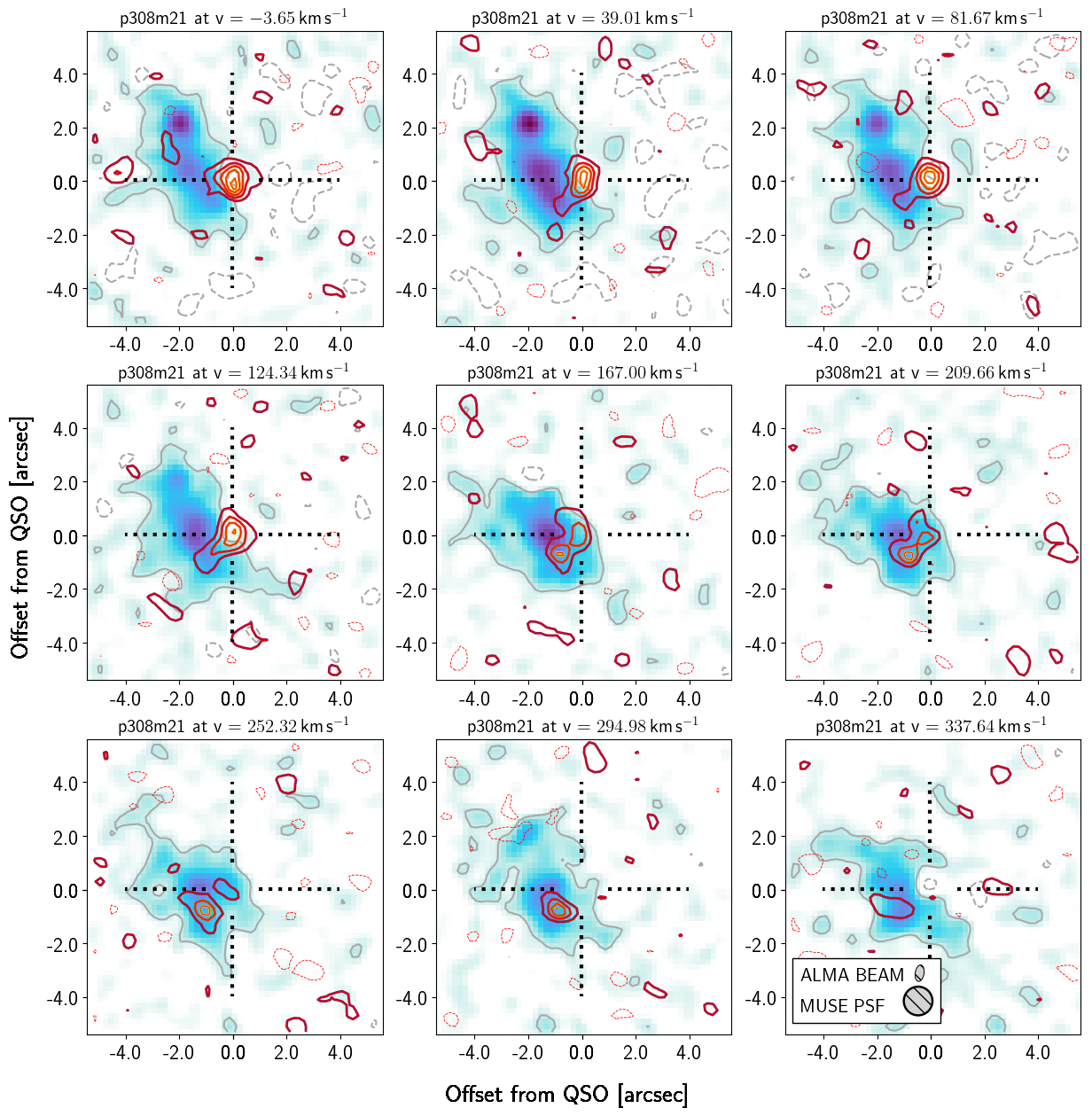}
\caption{\pmerg\ - continued from previous page}
\end{figure*}

\begin{figure*}\ContinuedFloat
\includegraphics[width=0.99\textwidth]{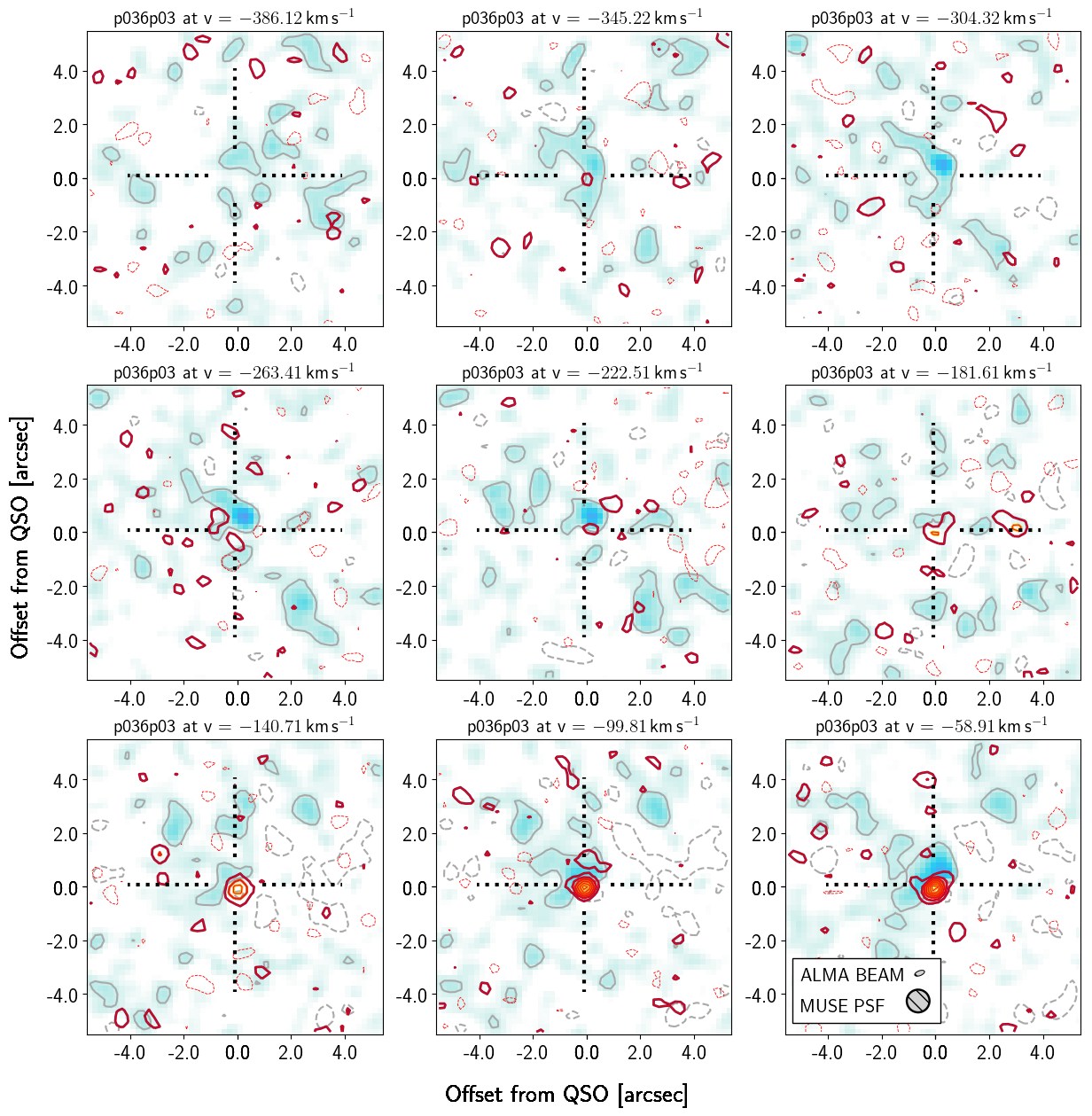}
\caption{\pthreesix}
\end{figure*}

\begin{figure*}\ContinuedFloat
\includegraphics[width=0.99\textwidth]{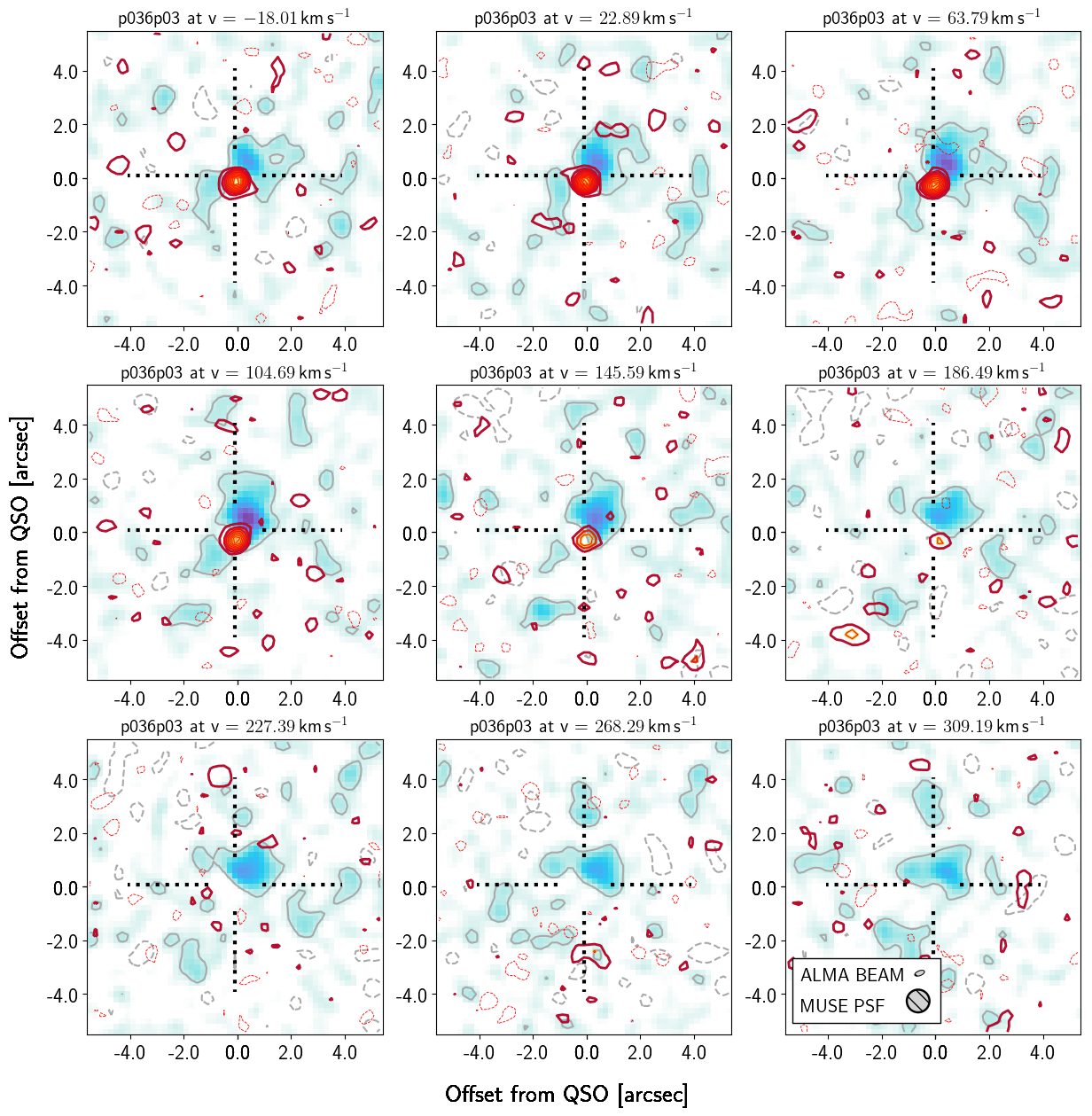}
\caption{\pthreesix\ - continued from previous page}
\end{figure*}

\begin{figure*}\ContinuedFloat
\includegraphics[width=0.99\textwidth]{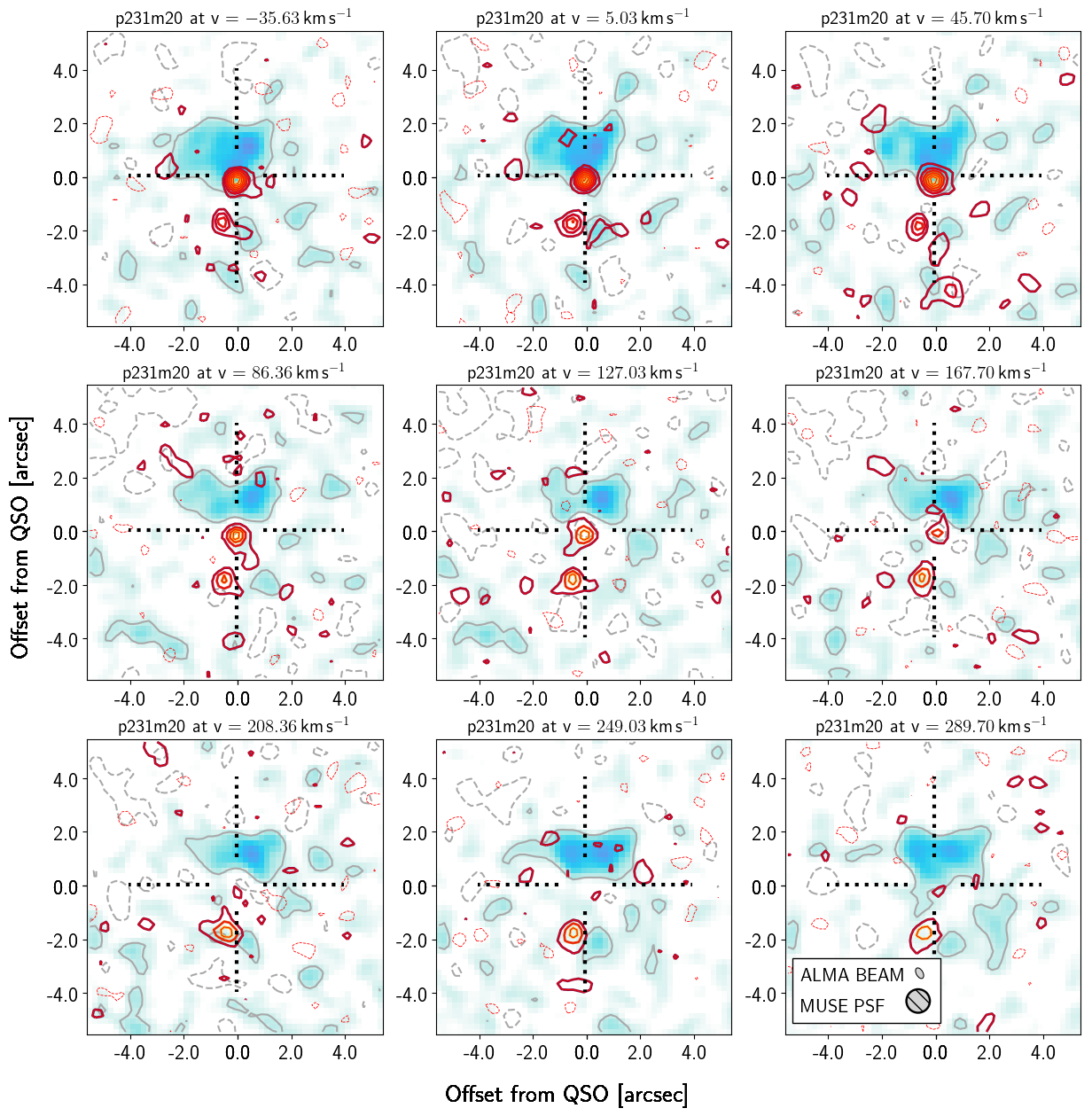}
\caption{\pchiara}
\end{figure*}

\begin{figure*}\ContinuedFloat
\includegraphics[width=0.99\textwidth]{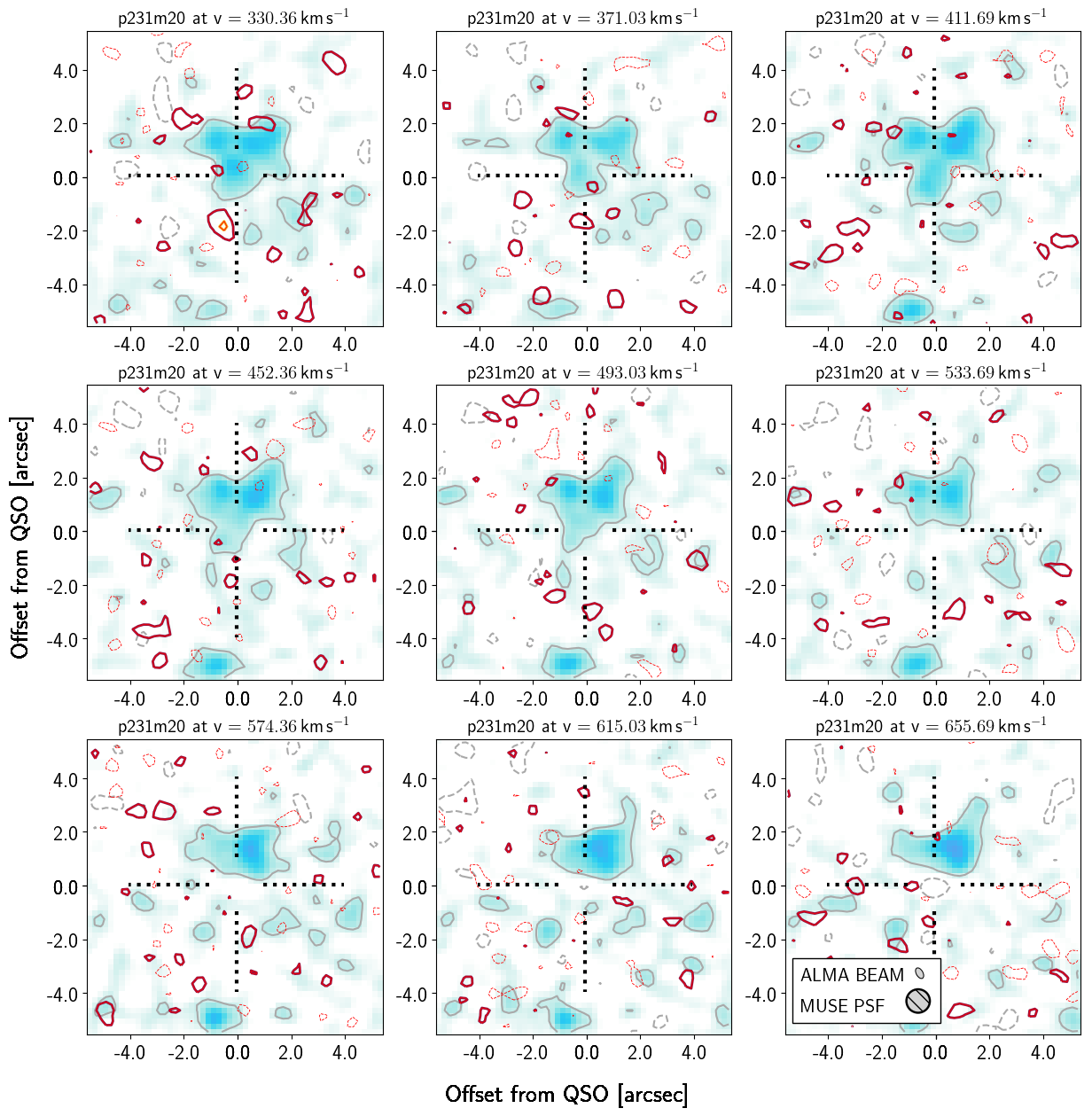}
\caption{\pchiara\ - continued from previous page}
\end{figure*}

\begin{figure*}\ContinuedFloat
\includegraphics[width=0.99\textwidth]{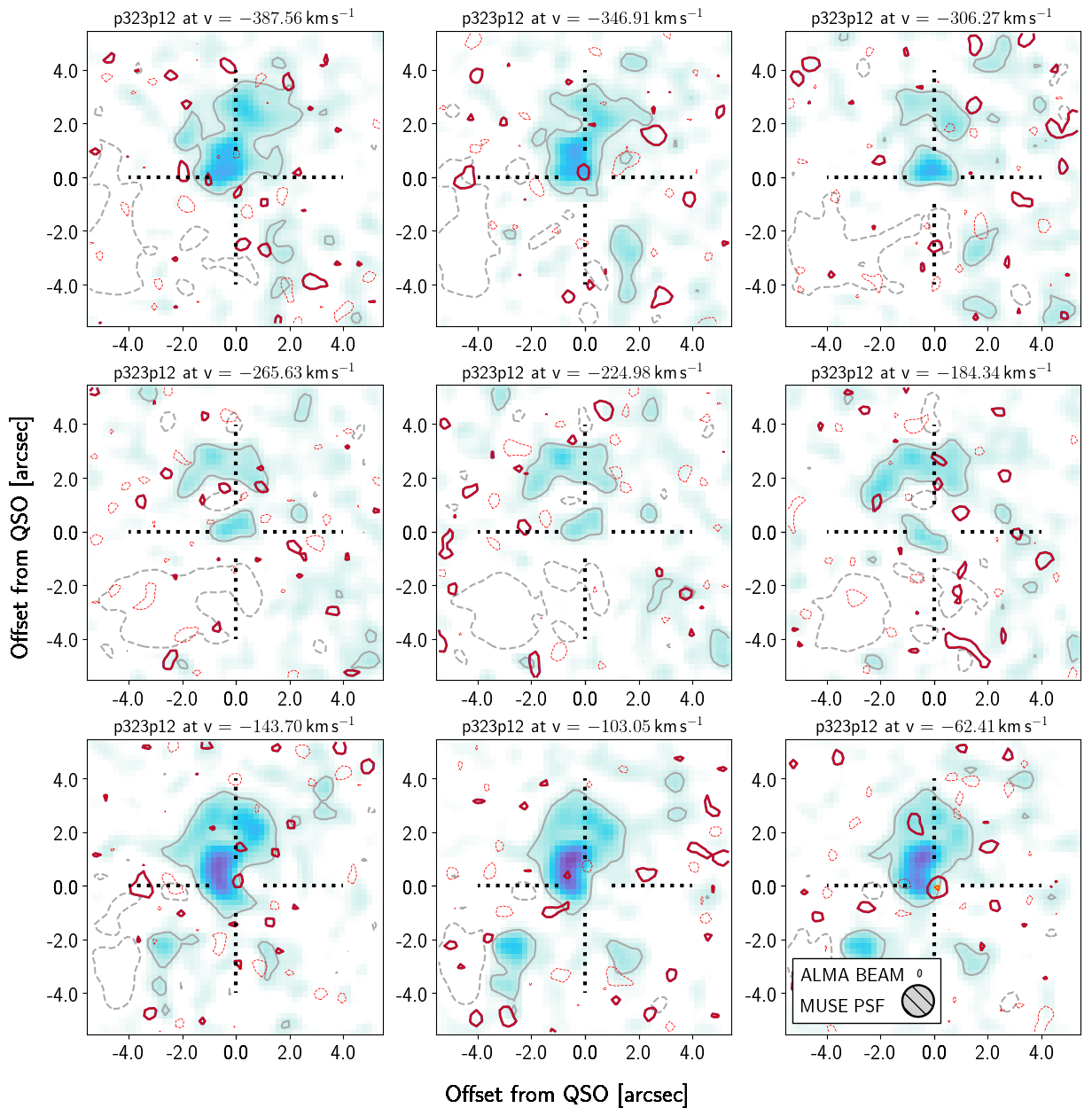}
\caption{\pthreetothree}
\end{figure*}

\begin{figure*}\ContinuedFloat
\includegraphics[width=0.99\textwidth]{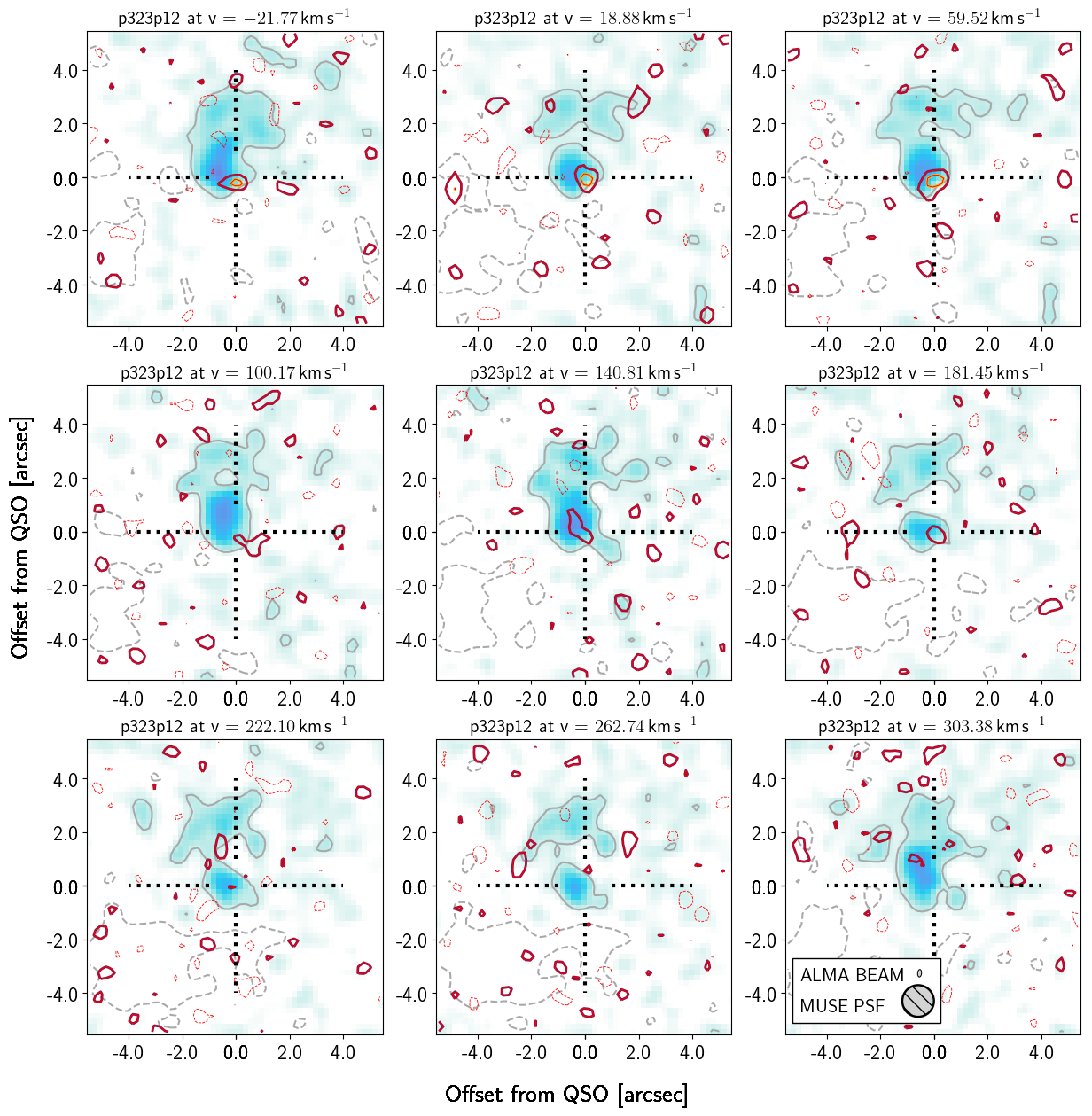}
\caption{\pthreetothree\ - continued from previous page}
\end{figure*}

\begin{figure*}\ContinuedFloat
\includegraphics[width=0.99\textwidth]{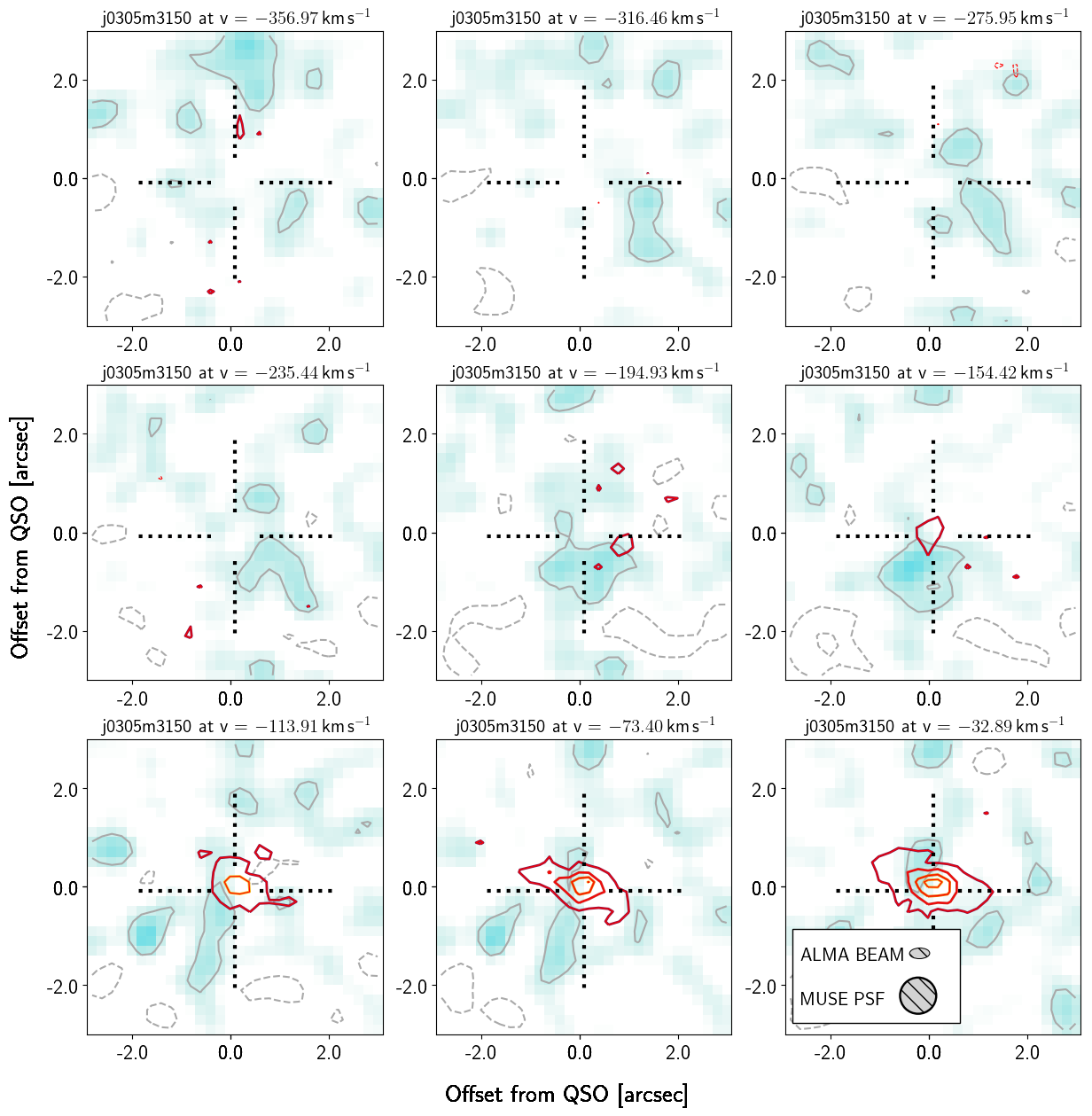}
\caption{\jema}
\end{figure*}

\begin{figure*}\ContinuedFloat
\includegraphics[width=0.99\textwidth]{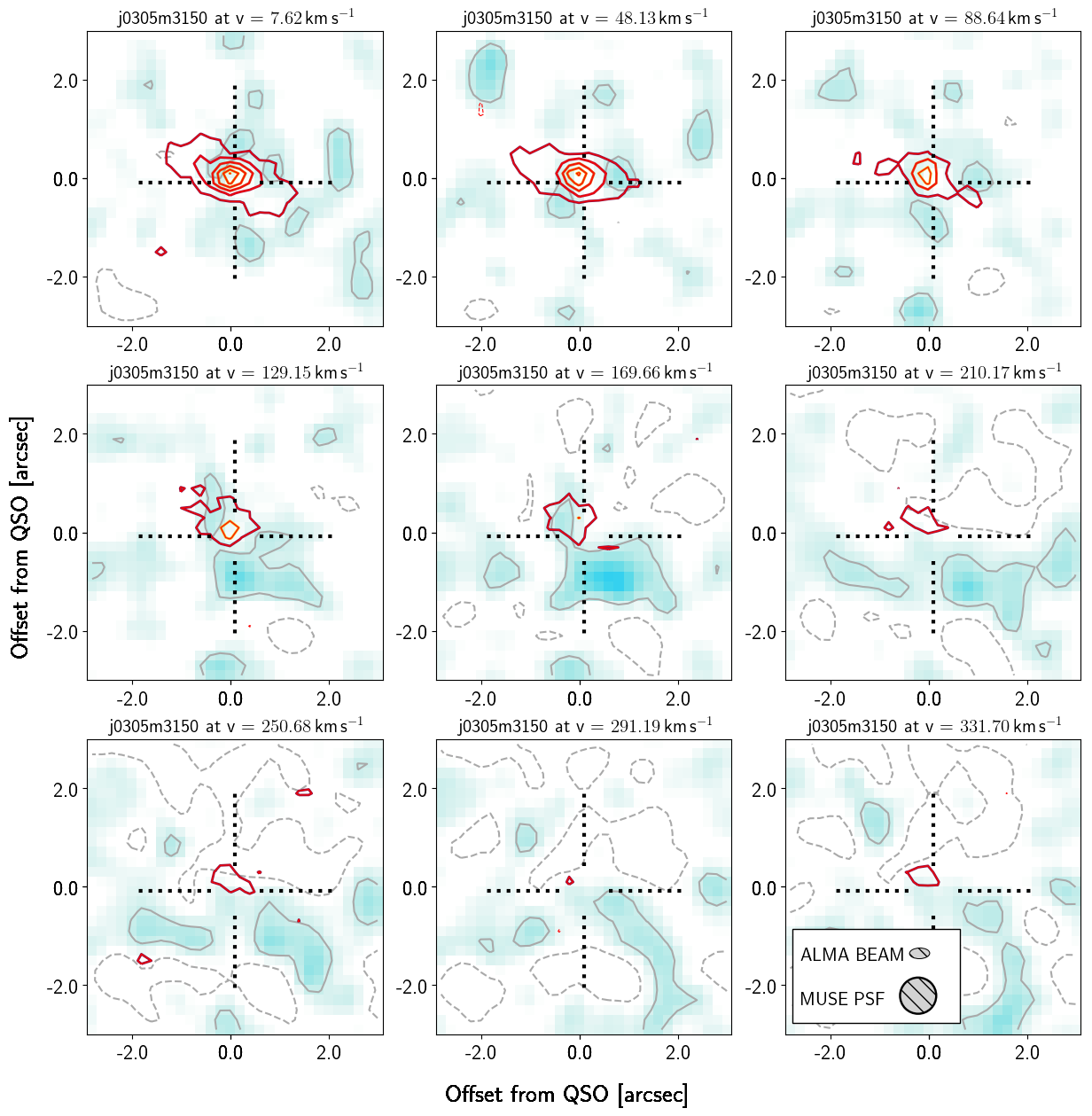}
\caption{\jema\ - continued from previous page}
\end{figure*}

\acknowledgments
We would like to thank the anonymous referee for their very constructive comments which have greatly helped to improve the paper. ABD, FW, MN and MN acknowledge support from the ERC Advanced Grant 740246 (Cosmic Gas). ABD also acknowledges support from the UK Science and Technology Facilities Council (STFC) under grant ST/V000624/1. \\

\bibliography{bib} 

\begin{thebibliography}{}
\expandafter\ifx\csname natexlab\endcsname\relax\def\natexlab#1{#1}\fi
\providecommand{\url}[1]{\href{#1}{#1}}

\bibitem[{{Arrigoni Battaia} {et~al.}(2019){Arrigoni Battaia}, Hennawi,
  Prochaska, O{\~{n}}orbe, Farina, Cantalupo, \& Lusso}]{ArrigoniBattaia2018}
{Arrigoni Battaia}, F., Hennawi, J.~F., Prochaska, J.~X., {et~al.} 2019, MNRAS,
  482, 3162

\bibitem[{Ba{\~{n}}ados {et~al.}(2016)Ba{\~{n}}ados, Venemans, Decarli, Farina,
  Mazzucchelli, Walter, Fan, Stern, Schlafly, Chambers, Rix, Jiang, McGreer,
  Simcoe, Wang, Yang, Morganson, Rosa, Greiner, Balokovi{\'{c}}, Burgett,
  Cooper, Draper, Flewelling, Hodapp, Jun, Kaiser, Kudritzki, Magnier,
  Metcalfe, Miller, Schindler, Tonry, Wainscoat, Waters, \& Yang}]{Banados2016}
Ba{\~{n}}ados, E., Venemans, B.~P., Decarli, R., {et~al.} 2016, The
  Astrophysical Journal Supplement Series, 227, 11

\bibitem[{Ba{\~{n}}ados {et~al.}(2018)Ba{\~{n}}ados, Venemans, Mazzucchelli,
  Farina, Walter, Wang, Decarli, Stern, Fan, Davies, Hennawi, Simcoe, Turner,
  Rix, Yang, Kelson, Rudie, \& Winters}]{Banados2018}
Ba{\~{n}}ados, E., Venemans, B.~P., Mazzucchelli, C., {et~al.} 2018, Nature,
  553, 473

\bibitem[{Borisova {et~al.}(2016)Borisova, Cantalupo, Lilly, Marino, Gallego,
  Bacon, Blaizot, Bouch{\'{e}}, Brinchmann, Carollo, Caruana, Finley, Herenz,
  Richard, Schaye, Straka, Turner, Urrutia, Verhamme, \&
  Wisotzki}]{Borisova2016}
Borisova, E., Cantalupo, S., Lilly, S.~J., {et~al.} 2016, arXiv:1605.01422

\bibitem[{Decarli {et~al.}(2017)Decarli, Walter, Venemans, Ba{\~{n}}ados,
  Bertoldi, Carilli, Fan, Farina, Mazzucchelli, Riechers, Rix, Strauss, Wang,
  \& Yang}]{Decarli2018}
Decarli, R., Walter, F., Venemans, B.~P., {et~al.} 2017, Nature Publishing
  Group, 545, doi:10.1038/nature22358

\bibitem[{{Decarli} {et~al.}(2019){Decarli}, {Dotti}, {Ba{\~n}ados}, {Farina},
  {Walter}, {Carilli}, {Fan}, {Mazzucchelli}, {Neeleman}, {Novak}, {Riechers},
  {Strauss}, {Venemans}, {Yang}, \& {Wang}}]{Decarli2019}
{Decarli}, R., {Dotti}, M., {Ba{\~n}ados}, E., {et~al.} 2019, \apj, 880, 157

\bibitem[{{Drake} {et~al.}(2019){Drake}, {Farina}, {Neeleman}, {Walter},
  {Venemans}, {Banados}, {Mazzucchelli}, \& {Decarli}}]{Drake2019}
{Drake}, A.~B., {Farina}, E.~P., {Neeleman}, M., {et~al.} 2019, \apj, 881, 131

\bibitem[{{Drake} {et~al.}(2020){Drake}, {Walter}, {Novak}, {Farina},
  {Neeleman}, {Riechers}, {Carilli}, {Decarli}, {Mazzucchelli}, \&
  {Onoue}}]{Drake20}
{Drake}, A.~B., {Walter}, F., {Novak}, M., {et~al.} 2020, arXiv e-prints,
  arXiv:2007.14221

\bibitem[{Farina {et~al.}(2017)Farina, Venemans, Decarli, Hennawi, Walter,
  Ba{\~{n}}ados, Mazzucchelli, Cantalupo, Arrigoni-Battaia, \&
  McGreer}]{Farina2017}
Farina, E.~P., Venemans, B.~P., Decarli, R., {et~al.} 2017, The Astrophysical
  Journal, 848, 78

\bibitem[{{Farina} {et~al.}(2019){Farina}, {Arrigoni-Battaia}, {Costa},
  {Walter}, {Hennawi}, {Drake}, {Decarli}, {Gutcke}, {Mazzucchelli},
  {Neeleman}, {Georgiev}, {Eilers}, {Davies}, {Ba{\~n}ados}, {Fan}, {Onoue},
  {Schindler}, {Venemans}, {Wang}, {Yang}, {Rabien}, \& {Busoni}}]{Farina2019}
{Farina}, E.~P., {Arrigoni-Battaia}, F., {Costa}, T., {et~al.} 2019, \apj, 887,
  196

\bibitem[{Goto {et~al.}(2009)Goto, Utsumi, Furusawa, Miyazaki, \&
  Komiyama}]{Goto2009}
Goto, T., Utsumi, Y., Furusawa, H., Miyazaki, S., \& Komiyama, Y. 2009, Mon.
  Not. R. Astron. Soc, 400, 843

\bibitem[{Goto {et~al.}(2012)Goto, Utsumi, Walsh, Hattori, Miyazaki, \&
  Yamauchi}]{Goto2012}
Goto, T., Utsumi, Y., Walsh, J.~R., {et~al.} 2012, Mon. Not. R. Astron. Soc,
  421, 77

\bibitem[{{Jiang} {et~al.}(2016){Jiang}, {McGreer}, {Fan}, {Strauss},
  {Ba{\~n}ados}, {Becker}, {Bian}, {Farnsworth}, {Shen}, {Wang}, {Wang},
  {Wang}, {White}, {Wu}, {Wu}, {Yang}, \& {Yang}}]{Jiang2016}
{Jiang}, L., {McGreer}, I.~D., {Fan}, X., {et~al.} 2016, \apj, 833, 222

\bibitem[{Mazzucchelli {et~al.}(2017)Mazzucchelli, Ba{\~{n}}ados, Venemans,
  Decarli, Farina, Walter, Eilers, Rix, Simcoe, Stern, Fan, Schlafly, {De
  Rosa}, Hennawi, Chambers, Greiner, Burgett, Draper, Kaiser, Kudritzki,
  Magnier, Metcalfe, Waters, \& Wainscoat}]{Mazzucchelli2017b}
Mazzucchelli, C., Ba{\~{n}}ados, E., Venemans, B.~P., {et~al.} 2017,
  arXiv:1710.01251

\bibitem[{Momose {et~al.}(2018)Momose, Goto, Utsumi, Hashimoto, Chiang, Kim,
  Kashikawa, Shimasaku, \& Miyazaki}]{Momose2018}
Momose, R., Goto, T., Utsumi, Y., {et~al.} 2018, MNRAS, 000, arXiv:1809.10916v1

\bibitem[{{Mortlock} {et~al.}(2011){Mortlock}, {Warren}, {Venemans}, {Patel},
  {Hewett}, {McMahon}, {Simpson}, {Theuns}, {Gonz{\'a}les-Solares}, {Adamson},
  {Dye}, {Hambly}, {Hirst}, {Irwin}, {Kuiper}, {Lawrence}, \&
  {R{\"o}ttgering}}]{Mortlock2011}
{Mortlock}, D.~J., {Warren}, S.~J., {Venemans}, B.~P., {et~al.} 2011, \nat,
  474, 616

\bibitem[{{Neeleman} {et~al.}(2019){Neeleman}, {Ba{\~n}ados}, {Walter},
  {Decarli}, {Venemans}, {Carilli}, {Fan}, {Farina}, {Mazzucchelli}, {Novak},
  {Riechers}, {Rix}, \& {Wang}}]{Neeleman2019}
{Neeleman}, M., {Ba{\~n}ados}, E., {Walter}, F., {et~al.} 2019, \apj, 882, 10

\bibitem[{{Neeleman} {et~al.}(2021){Neeleman}, {Novak}, {Venemans}, {Walter},
  {Decarli}, {Kaasinen}, {Schindler}, {Banados}, {Carilli}, {Drake}, {Fan}, \&
  {Rix}}]{Neeleman2021}
{Neeleman}, M., {Novak}, M., {Venemans}, B.~P., {et~al.} 2021, arXiv e-prints,
  arXiv:2102.05679

\bibitem[{{Novak} {et~al.}(2020){Novak}, {Venemans}, {Walter}, {Neeleman},
  {Kaasinen}, {Liang}, {Feldmann}, {Ba{\~n}ados}, {Carilli}, {Decarli},
  {Drake}, {Fan}, {Farina}, {Mazzucchelli}, {Rix}, \& {Wang}}]{Novak20}
{Novak}, M., {Venemans}, B.~P., {Walter}, F., {et~al.} 2020, \apj, 904, 131

\bibitem[{{Prescott} {et~al.}(2015){Prescott}, {Martin}, \& {Dey}}]{Prescott15}
{Prescott}, M. K.~M., {Martin}, C.~L., \& {Dey}, A. 2015, \apj, 799, 62

\bibitem[{Roche {et~al.}(2014)Roche, Humphrey, \& Binette}]{Roche2014}
Roche, N., Humphrey, A., \& Binette, L. 2014, MNRAS, 443, 3795

\bibitem[{{Schindler} {et~al.}(2020){Schindler}, {Farina}, {Ba{\~n}ados},
  {Eilers}, {Hennawi}, {Onoue}, {Venemans}, {Walter}, {Wang}, {Davies},
  {Decarli}, {Rosa}, {Drake}, {Fan}, {Mazzucchelli}, {Rix}, {Worseck}, \&
  {Yang}}]{Schindler2021}
{Schindler}, J.-T., {Farina}, E.~P., {Ba{\~n}ados}, E., {et~al.} 2020, \apj,
  905, 51

\bibitem[{Stewart {et~al.}(2013)Stewart, Brooks, Bullock, Maller, Diemand,
  Wadsley, \& Moustakas}]{Stewart2013}
Stewart, K.~R., Brooks, A.~M., Bullock, J.~S., {et~al.} 2013, The Astrophysical
  Journal, 769, 74.
\newblock \url{https://doi.org/10.1088%2F0004-637x%2F769%2F1%2F74}

\bibitem[{Stewart {et~al.}(2011)Stewart, Kaufmann, Bullock, Barton, Maller,
  Diemand, \& Wadsley}]{Stewart2011}
Stewart, K.~R., Kaufmann, T., Bullock, J.~S., {et~al.} 2011, The Astrophysical
  Journal, 738, 39.
\newblock \url{https://doi.org/10.1088%2F0004-637x%2F738%2F1%2F39}

\bibitem[{{Venemans} {et~al.}(2019){Venemans}, {Neeleman}, {Walter}, {Novak},
  {Decarli}, {Hennawi}, \& {Rix}}]{Venemans2019}
{Venemans}, B.~P., {Neeleman}, M., {Walter}, F., {et~al.} 2019, \apjl, 874, L30

\bibitem[{{Venemans} {et~al.}(2016){Venemans}, {Walter}, {Zschaechner},
  {Decarli}, {De Rosa}, {Findlay}, {McMahon}, \& {Sutherland}}]{venemans2016}
{Venemans}, B.~P., {Walter}, F., {Zschaechner}, L., {et~al.} 2016, \apj, 816,
  37

\bibitem[{{Venemans} {et~al.}(2020){Venemans}, {Walter}, {Neeleman}, {Novak},
  {Otter}, {Decarli}, {Ba{\~n}ados}, {Drake}, {Farina}, {Kaasinen},
  {Mazzucchelli}, {Carilli}, {Fan}, {Rix}, \& {Wang}}]{Venemans20}
{Venemans}, B.~P., {Walter}, F., {Neeleman}, M., {et~al.} 2020, \apj, 904, 130

\bibitem[{Willott {et~al.}(2011)Willott, Chet, Bergeron, \&
  Hutchings}]{Willott2011}
Willott, C.~J., Chet, S., Bergeron, J., \& Hutchings, J.~B. 2011, The
  Astronomical Journal, 142, arXiv:1109.4110v1

\bibitem[{{Yang} {et~al.}(2020){Yang}, {Wang}, {Fan}, {Hennawi}, {Davies},
  {Yue}, {Banados}, {Wu}, {Venemans}, {Barth}, {Bian}, {Boutsia}, {Decarli},
  {Farina}, {Green}, {Jiang}, {Li}, {Mazzucchelli}, \& {Walter}}]{Yang2020}
{Yang}, J., {Wang}, F., {Fan}, X., {et~al.} 2020, \apjl, 897, L14

\bibitem[{Zeimann {et~al.}(2011)Zeimann, {Richard L. White}, {Robert H.
  Becker}, {Jacqueline A. Hodge}, {Spencer A. Stanford}, \& {Gordon T.
  Richards}}]{Zeimann2011}
Zeimann, G., {Richard L. White}, {Robert H. Becker}, {et~al.} 2011, The
  Astrophysical Journal,, 736

\end{thebibliography}

\end{document}